\documentclass[showpacs,twocolumn,superscriptaddress]{revtex4-1}
\usepackage{doi}
\usepackage{hyperref}
\hypersetup{
  colorlinks=true,        
  linkcolor=blue,         
  citecolor=cyan,         
}

\usepackage{graphicx}
\usepackage{dcolumn}
\usepackage{bm}
\usepackage{xcolor, soul}
\usepackage{color}
\usepackage[normalem]{ulem}  
\usepackage{enumitem}
\usepackage{amssymb}
\usepackage{orcidlink}
\usepackage{graphicx} 
\usepackage{makecell}

\usepackage{amsmath}
\usepackage{cleveref}

\crefname{section}{Sec.}{Secs.}
\Crefname{section}{Sec.}{Secs.}
\crefname{equation}{Eq.}{Eqs.}
\Crefname{equation}{Eq.}{Eqs.}   

\begin{document}

\title{Magnetic Modification of Black Hole Photospheres with Image Contraction, Efficiency Shifts and Redshift Boosts in Schwarzschild-Bertotti-Robinson Spacetime}

\author{Javokhir Sharipov}
\email{javohirsh100@gmail.com}
\affiliation{Institute of Fundamental and Applied Research, National Research University TIIAME, Kori Niyoziy 39, Tashkent 100000, Uzbekistan}

\author{Pankaj Sheoran
}
\email{pankaj.sheoran@vit.ac.in}
\affiliation{School of Advanced Sciences, Vellore Institute of Technology, Tiruvalam Rd, Katpadi, Vellore, Tamil Nadu 632014, India}

\author{Sanjar Shaymatov}
\email{sanjar@astrin.uz}
\affiliation{Institute of Fundamental and Applied Research, National Research University TIIAME, Kori Niyoziy 39, Tashkent 100000, Uzbekistan}
\affiliation{University of Tashkent for Applied Sciences, Str. Gavhar 1, Tashkent 100149, Uzbekistan}
\affiliation{Tashkent State Technical University, 100095 Tashkent, Uzbekistan}

\date{\today}

\begin{abstract}
{We investigate the optical and radiative signatures of an accretion disk around a Schwarzschild black hole (BH) immersed in a uniform magnetic field. The spacetime geometry is described by the Schwarzschild-Bertotti-Robinson (SBR) metric, which represents the non-rotating sector of the recently discovered Kerr-Bertotti-Robinson exact solution to the Einstein-Maxwell equations. We begin with the study of null geodesics and demonstrate that the self-consistent magnetic field fundamentally alters photon propagation, causing an expansion of light bundles relative to the Schwarzschild case due to modified initial conditions in the orbital equation. We then compute the magnetic field-dependent shifts of key characteristic radii: the event horizon ($r_h$), photon sphere ($r_{ph}$), and innermost stable circular orbit ($r_{ISCO}$). We find that all three increase monotonically with field strength $B$, revealing a magnetic amplification of the effective gravitational field. For $B=0.05$, we find that the lensed emission bands contract to a narrower impact parameter range, $b\in(4.976,5.149)\cup(5.19,6.128)$. Employing ray-tracing formalism, we construct observed accretion disk images and quantify magnetic modifications, showing that the direct image contracts while maximum energy flux, radiation temperature, and redshift factor are enhanced. Complementing these numerical findings, we develop an analytical framework for the accretion disk dynamics. We derive the modified Keplerian frequency $\Omega_K$, along with the specific energy $E$ and angular momentum $L$ for circular orbits. From these, we obtain the exact ISCO radius $r_{\text{ISCO}}$, which shows an outward shift. This outward shift reveals that the radiative efficiency decreases dramatically with increasing magnetic field strength $B$. For $\beta = BM \sim 0.1$, the efficiency drops by approximately $91\%$. Finally, we conduct a redshift analysis, and interestingly, at an inclination $\theta_0=80^\circ$ and $B=0.05$, we obtain $z_{\max}=1.17$ and $z_{\min}=-0.3$, compared to $z_{\max}=1.11$ and $z_{\min}=-0.28$ for the Schwarzschild case, providing a distinct and testable signature. These magnetic imprints provide observational discriminants for probing strong-field regimes in BH-magnetar binaries and establish an essential baseline for understanding how self-consistent electromagnetic fields modify BH observables.}
 
\end{abstract}

\maketitle

\section{Introduction}

Einstein's General Relativity (GR) \cite{1983mtbh.book.....C} revolutionized our understanding of gravity by describing it as a geometric property of spacetime. Among the most profound predictions of the field equations are black holes (BHs) \cite{Bambi:2015kza, Bambi:2017khi}, regions of spacetime exhibiting such strong gravitational effects that nothing, not even light, can escape. For decades, these objects remained purely theoretical. However, this changed dramatically with recent astronomical breakthroughs. The first direct image of a BH's shadow, captured by the Event Horizon Telescope (EHT) collaboration in the galaxy M87* \cite{Vagnozzi23EHT}, and subsequently of Sagittarius A$^*$ (Sgr A$^*$) at the center of our own Milky Way \cite{EventHorizonTelescope:2022wkp, Ghez05,Nucita07}, provided not only stunning visual confirmation but also opened a new era of precision tests for gravity in its strongest regimes.

Astrophysically, BHs are not merely exotic endpoints of stellar evolution; they are fundamental engines powering some of the most luminous and energetic phenomena in the universe. They are central to the understanding of quasars \cite{Morgan10ApJ,Czerny23ApSS}, X-ray binaries \cite{1974ARA&A..12...23B,Yao:2001xq,Bu23BOOK,2006csxs.book.....L}, and active galactic nuclei \cite{2017A&ARv..25....2P,2024A&AT...34..249S}. The accretion of matter onto these compact objects is an extraordinarily efficient process, converting gravitational potential energy into radiation as the plasma spirals inward. This makes the accretion disk one of the best natural laboratories for probing the dynamical properties of particles near the BH and for testing the predictions of GR theories against observational data, such as X-ray radiation spectra and continuum fitting methods \cite{Abramowicz:2011xu,Fender:2004gg,Bambi:2012ku, Parker:2019ryb,Wen:2020cpm}. The detailed study of these disks allows us to infer the spacetime geometry and constrain alternative gravity models \cite{Perez:2012bx, Perez:2017spz, Kazempour:2022asl,Wu:2024sng,Soroushfar:2020kgb}.

The recent dawn of gravitational wave astronomy has further elevated the importance of BHs as laboratories for fundamental physics. The LIGO-Virgo-KAGRA collaboration's direct detection of gravitational waves from merging BHs \cite{LIGOScientific:2016aoc, KAGRA:2021vkt} has not only confirmed a major prediction of GR but also opened a completely new observational window. These detections allow for unprecedented tests of strong-field gravity, including constraints on BH ringdown modes and tests of the no-hair theorem \cite{Isi:2019aib, LIGOScientific:2020tif}. Future gravitational wave observatories, such as the Laser Interferometer Space Antenna (LISA) \cite{2017arXiv170200786A}, will be sensitive to extreme mass-ratio inspirals (EMRIs), systems where a compact object spirals into a supermassive BH. EMRIs are exquisite probes of the spacetime geometry, as their gravitational waveforms encode detailed information about the central BH's mass, spin, and any deviations from the Kerr metric due to surrounding fields or modifications to gravity \cite{Gair:2012nm, Babak:2017tow}. Studying how electromagnetic fields, like those in the Bertotti-Robinson spacetime, imprint on these waveforms is therefore of paramount importance for future missions \cite{Li:2025rtf}.

In this context, a particularly interesting family of spacetimes is the Bertotti-Robinson (BR) class \cite{Bertotti:1959pf,Robinson:1959ev}, which describes a homogeneous electromagnetic universe with an $\text{AdS}_2 \times S^2$ geometry. While much attention has recently been given to the rotating generalization, the Kerr-Bertotti-Robinson (KBR) metric \cite{Podolsky:2025tle,Ovcharenko:2025cpm}, which belongs to the larger Pleba\'nski-Demia\'nski class \cite{1975AnPhy..90..196P, Plebanski:1976gy,PhysRevLett.37.493} and was first systematically investigated by Carter \cite{Carter:1968ks}, our focus in this work is on its non-rotating counterpart. 

At first glance, focusing on a non-rotating BH might seem restrictive, since most astrophysical BHs are expected to possess some angular momentum. However, the non-rotating Bertotti-Robinson BH serves as a crucial and theoretically clean laboratory for several compelling reasons. First, it provides a non-perturbative, exact solution to the Einstein-Maxwell equations where a uniform magnetic field \cite{Frolov10,Aliev02,Shaymatov14,Tursunov16,Shaymatov22a,Pavlovic19,Shaymatov19b,Duztas-Jamil20,Shaymatov20egb} is treated as a self-consistent part of the spacetime geometry, rather than a perturbation on a Kerr background. This framework is essential for understanding environments where magnetic and gravitational energy densities are comparable, such as in magnetically dominated accretion flows \cite{Bocquet:1995je} or in binary systems composed of a BH and a magnetar \cite{Ginzburg1964,Rezzolla01,deFelice03}. 
Second, and equally important, is the question of accretion itself. A non-rotating BH can indeed possess an accretion disk. While rotation certainly modifies the disk structure through frame-dragging effects, the fundamental mechanism of accretion, matter losing angular momentum and spiraling inward via viscous processes (e.g., turbulence driven by the magneto-rotational instability (MRI) or shock waves), operates independently of the BH's spin.
Spherically symmetric (Schwarzschild) BHs have been the foundation of accretion disk theory since the pioneering work of Novikov and Thorne \cite{1973blho.conf..343N} and Shakura and Sunyaev \cite{Shakura:1972te}. In fact, many of the key features of accretion disks \cite{Bambi17e,Faraji:2025,Boos:2025nzc}, such as the existence of an innermost stable circular orbit (ISCO) \cite{Dadhich22a,Dadhich22IJMPD}, the radial profiles of temperature and flux, and the overall spectral energy distribution \cite{Gyulchev20,Boshkayev:2020kle,Gyulchev2021E,Shaymatov2023,Collodel_2021,Alloqulov24CPC,Boshkayev21PRD,Alloqulov24EPJP,Liu:2020ola,Nozari2025,Liu:2020vkh,Igata2025,Nozari2025b,Zhu:2019ura,Rehman:2026vxm,Jiang:2023img,Yan:2025mlg}, are qualitatively similar for both rotating and non-rotating BHs, with spin primarily causing quantitative shifts (e.g., ISCO location, efficiency). Therefore, studying the non-rotating case allows us to isolate and cleanly identify the specific imprints of the magnetic field on disk observables, disentangling them from the already complex effects of rotation.

From a methodological perspective, exact solutions like the Bertotti-Robinson metric are invaluable for developing and testing numerical relativity codes and analytical approximation methods. They serve as controlled backgrounds against which more complicated, realistic scenarios can be benchmarked. Furthermore, the study of such spacetimes bridges the gap between pure theory and observation. 
By calculating observable quantities, such as accretion disk spectra, polarization signatures, and gravitational wave forms, from first principles in an exact spacetime. We can build robust templates for searching for magnetic field effects in real astrophysical data.
This is particularly relevant for upcoming X-ray polarimetry missions \cite{2016SPIE.9905E..17W, 2019SPIE11118E..0VO} and next-generation gravitational wave detectors, which promise to deliver data of sufficient quality to probe the electromagnetic environment around BHs.

It is also important to recognize that many of the fundamental physical processes we investigate in the non-rotating BR spacetime will remain qualitatively, and in some cases quantitatively, relevant when extended to the rotating KBR case.
For instance, the magnetic field modifies particle orbits through the combined effects of gravitational and Lorentz forces on charged particles \cite{Shaymatov15,Shaymatov21pdu}. These modifications will persist in the rotating spacetime, although they acquire additional complexity from frame-dragging.
The general behavior of the redshift factor, which combines gravitational and Doppler contributions, will share common features, though rotation will introduce additional angular dependencies. The formation of accretion disk images via ray tracing depends on the underlying geodesic structure \cite{Cui24,Chen2025,Ziqiang25,Liu:2021yev,Sharipov:2025yfw,Hu2023}, and the magnetic field's influence on photon trajectories (through the effective potential) will be analogous in both spacetimes. What will change are the quantitative details: the precise location of the ISCO, the specific energy and angular momentum profiles of the disk, and the resulting flux distribution. However, the foundational understanding gained from the non-rotating case, how a self-consistent magnetic field modifies emission patterns and observational signatures, provides an essential baseline for interpreting the more complex rotating scenarios. Thus, our work not only fills a specific gap but also establishes a framework for future extensions to KBR. 

Significant work has been conducted on the rotating KBR spacetime, exploring its rich phenomenology. Extensive studies have investigated photon motion, geodesics, and BH shadows, demonstrating how a uniform magnetic field alters trajectories and optical signatures compared to the standard Kerr case \cite{Wang:2025vsx,Vachher:2025jsq,Halilsoy:2024sac}. Further research has analyzed energy extraction mechanisms like magnetic reconnection and the magnetic Penrose process \cite{Zeng:2025olq}, and potential imprints on gravitational-wave signals \cite{Li:2025rtf}. However, a systematic and detailed analysis of the accretion disk properties, radiation spectra, and resulting observed images specifically for the \textit{non-rotating} Bertotti-Robinson BH immersed in a uniform magnetic field remains a gap in the literature. This work aims to fill that gap by investigating how this exact magnetic field solution affects the motion of test particles and, consequently, the structure and emission characteristics of its surrounding accretion disk.

Based on the accretion disk models \cite{1973blho.conf..343N,Narayan:1994et,Abramowicz:1988sp}, the properties of the accretion disk have been analyzed with modified gravitational theories \cite{Kazempour:2022asl,Wu:2024sng,Soroushfar:2020kgb}. Analyzing the accretion disk around a BH within a homogeneous magnetic field and its observed image is very important and attractive, such as in astrophysical binary systems composed of a magnetar and a BH. For a BH within the magnetic field, there are several solutions of the space-time metric \cite{Melvin:1963qx,ernst1976black,Hiscock:1980zf,gibbons2013,Shaymatov22PhRvD.106b4039S,Shaymatov22PhRvD.106b4039S,Shaymatov22EPJC...82..636S}. In this work, we test the novel space-time metric \cite{Podolsky:2025tle} and investigate how the magnetic field affects the motions of the geodesics and the properties of the accretion disk. Namely, redshift distribution, radiation properties, and observed images are analyzed. By this study, we can highlight the physical nature near the BH, which is crucial for prospective studies.

This work is structured as follows: In Sec.~\ref{Sec:II}, we present the space-time metric expression and photon motion near the BH. Then we devote Sec.~\ref{Sec:III} to the accretion disk orbiting a BH within a uniform magnetic field. This section is separated into subsections: Image formation theory (\ref{subsec:A}), radiation properties on the surface of the disk (\ref{subs:B}), redshift factor, and the observed image at infinity (\ref{subs:c}). Finally, we summarize the results and conclude in Sec.~\ref{Sec:conclusion}.

\section{Spacetime metric and null geodesics}\label{Sec:II}

In this section, we study the null geodesics around a Schwarzschild BH in a uniform Bertotti–Robinson magnetic field (hereafter, the Schwarzschild–Bertotti–Robinson (SBR) BH).
For this, we use the solution obtained by J.Podolsky and H.Ovcharenko, given below~\cite{Podolsky:2025tle}
\begin{eqnarray}\label{eq:metric}
ds^2 &=& \frac{1}{\Omega^2} \Bigg[
- \mathcal{Q} dt ^2
+ \frac{dr^2}{\mathcal{Q}} 
+ r^2\left(\frac{d\theta^2}{P} \, 
+ {P}\sin^2\theta   \, d\phi^2\right)
\Bigg]\, ,\nonumber\\ 
\end{eqnarray}
with 
\begin{eqnarray}
P &=& 1 + B^2 M^2 \cos^2\theta,\nonumber\\
\mathcal{Q} &=& (1 + B^2 r^2)\left(1-\frac{2M}{r}-B^2 M^2\right),\nonumber\\
\Omega^2 &=& 1 + B^2 \left[r^2\sin{\theta}^2+Mr(2+ B^2Mr) \cos^2\theta\right]\, ,
\end{eqnarray}
where $M$ and $B$ are the BH mass and the magnetic field strength, respectively. The spacetime metric simplifies to the Bertotti-Robinson spacetime in the limit when $M=0$, while to the Schwarzschild metric when $B=0$. 

By applying Lagrangian formalism, we now analyze the motion of massless particles around the  SBR BH~\cite{1983mtbh.book.....C,Misner:1973prb}

\begin{eqnarray}
\mathcal{L}=\frac{1}{2}m\,g_{\mu\nu}\,\frac{dx^{\mu}}{d\tau}\,\frac{dx^{\nu}}{d\tau},
\end{eqnarray}
where $\tau$ and $m$ represent the affine parameter and the rest mass of the particle, respectively. For simplicity, we set $m=1$. Then we obtain the generalized momenta of the particle, only two of them coincide with the conserved quantities (energy $E$ and angular momentum $L$):
\begin{eqnarray}
    p_{\mu}&=&\frac{\partial {\cal L}}{\partial \dot{x}^{\mu}}=g_{\mu \nu}\dot{x}^{\nu}\,,\\
\label{eq:eqmotion}
p_{t} &=& g_{tt}\dot{t} = -\frac{\mathcal{Q}}{\Omega^2} \dot{t}=-E \, , \\
p_{\phi} &=&g_{\phi\phi}\dot{\phi} = \frac{Pr^{2}\sin^{2}\theta}{\Omega^2}\dot{\phi}=L \label{eq:momentum} \, . 
\end{eqnarray}
\begin{figure*}
\centering
  \includegraphics[scale=0.39]{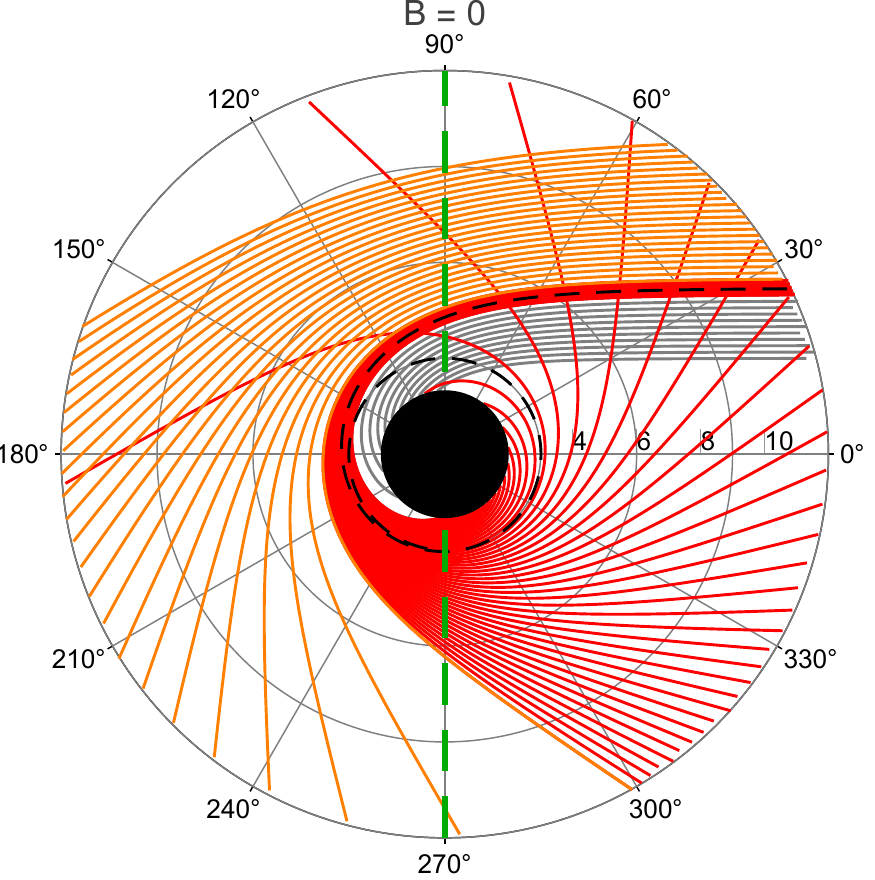}
  \includegraphics[scale=0.39]{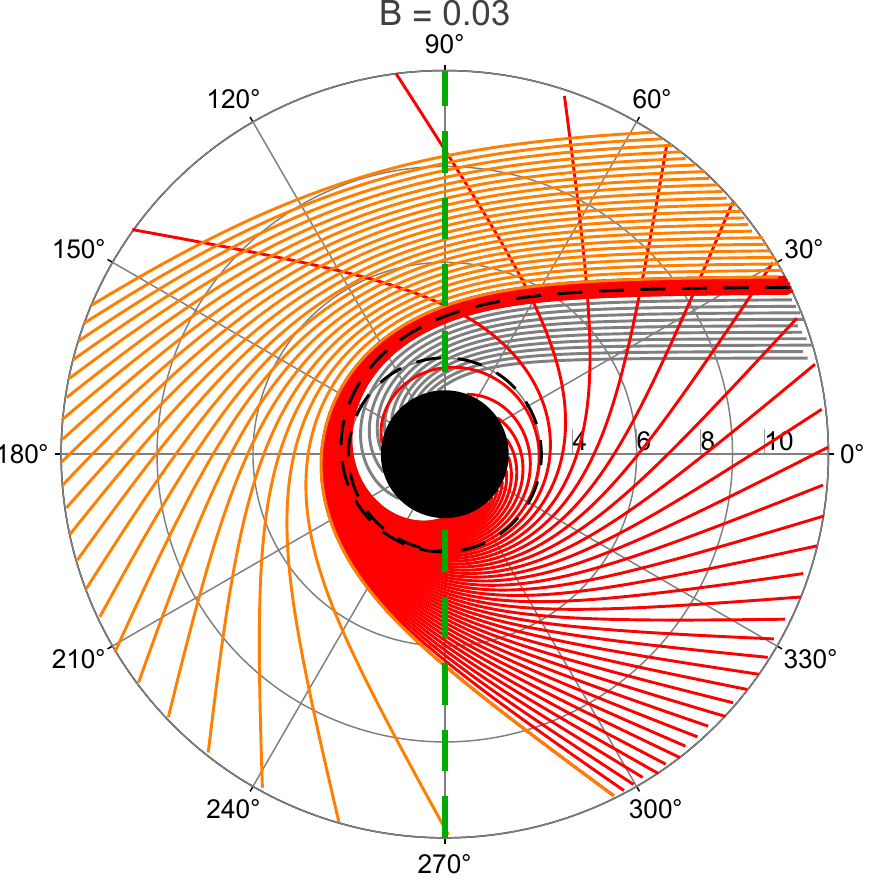}
  \includegraphics[scale=0.39]{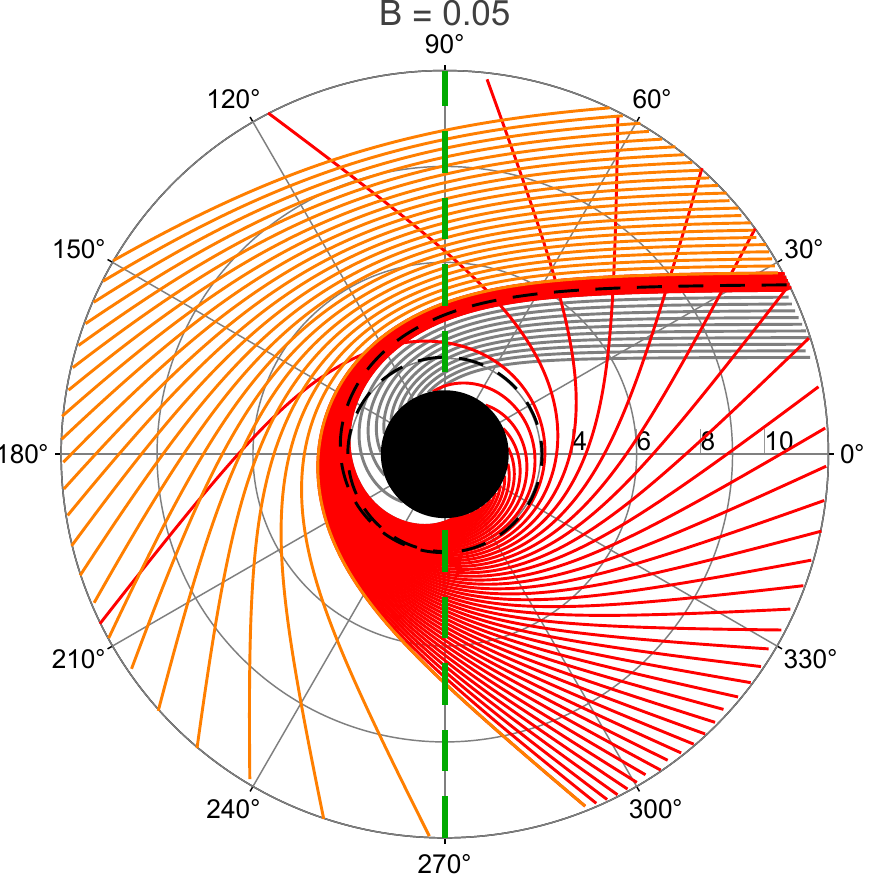}
  \caption{The plots demonstrate photon trajectories near the SBR BH for different values of $B$. The colored curves depict separate ranges of the impact parameter $b$: gray for $3<b\leq5$, red for $5<b\leq5.5$, and orange for $b>5.5$. The black dashed ring denotes the photon-sphere location.}
\label{fig:ray1}
\end{figure*}

We now reduce the above equations to a simpler form in the equatorial plane $(\theta=\pi/2)$ and substitute the resulting expression into the null geodesic equation $g_{\mu\nu}\dot{x}^\mu\dot{x}^\nu=0$ to obtain the radial equation for photons around the SBR BH.
\begin{eqnarray}
    \dot{r}^2=\Omega^4\Big[E^2-\mathcal{Q}\frac{L^2}{r^2}\Big]\ ,\label{eq:lamda}
\end{eqnarray}

\begin{table*}
\centering
\caption{Significant quantities for the SBR BH at different magnetic field $B$ values, along with impact parameter $b$ intervals separating the three categories of photon trajectories, are presented.}
\label{tab:nb}
\renewcommand{\arraystretch}{1.3}
\begin{tabular}{|c|c|c|c|c|c|c|c|c|}
\hline
\textbf{BHs} &
\textbf{$r_h$} & \textbf{$r_{ph}$} & \textbf{$b_c$} & $r_{ISCO}$ & \makecell{\textbf{direct emission}\\ $\pi/2\le\varphi<3\pi/2$} & \makecell{\textbf{lensed emission}\\$3\pi/2\le\varphi<5\pi/2$} & \makecell{\textbf{photon-ring}\\$\varphi\geq5\pi/2$}\\
\Xhline{1.3pt}

 \makecell{B=0 \\(Schw BH)} & 2& 3& 5.196 & 6 & $b\in(2.848,5.015)\cup(6.168,\infty)$ & $b\in(5.015,5.188)\cup(5.228,6.168)$ & $b\in(5.188,5.228)$ \\ \hline
 \textbf{$B=0.03$} & 2.002 & 3.011 & 5.182 & 6.005 & $b\in(2.846,5.001)\cup(6.153,\infty)$ & $b\in(5.001,5.174)\cup(5.214,6.153)$ & $b\in(5.174,5.214)$\\ \hline
 \textbf{$B=0.05$} & 2.005 & 3.031 & 5.157 & 6.015 & $b\in(2.842,4.976)\cup(6.128,\infty)$ & $b\in(4.976,5.149)\cup(5.19,6.128)$ & $b\in(5.149,5.19)$\\ \hline
 
\end{tabular}
\end{table*}
For convenience, we express the above equation in terms of the azimuthal angle $\phi$. To do this, we use the expression from Eq.~\eqref{eq:momentum} in the equatorial plane, rewrite $\dot{r}$, and then substitute it as follows
\begin{eqnarray}
    \dot{r}=\frac{dr}{d\phi}\frac{d\phi}{d\tau}=\frac{L\Omega^2}{r^2}\frac{dr}{d\phi}\, ,
\end{eqnarray}
\begin{eqnarray}
    \frac{1}{r^4}\left(\frac{dr}{d\phi}\right)^2=\frac{1}{b^2}-\frac{\mathcal{Q}}{r^2}=\tilde{V}_{eff}(r)\, ,
    \label{radialPhi}
\end{eqnarray}
where $b = L/E$, $\tilde{V}_{\mathrm{eff}}$ denotes the normalized effective potential. From condition $d\tilde{V}_{eff}/dr=0$, we obtain the radius of the photon sphere $r_{ph}$, and from condition $\tilde{V}_{eff}(r_{ph})=0$, we determine the critical impact parameters $b_c$~\cite{hartmann2023gr}:
\begin{eqnarray}
b_c = \frac{r_{ph}}{\sqrt{\mathcal{Q}(r_{ph})}}\, ,
\end{eqnarray}
where
\begin{eqnarray}
    r_{\rm ph}=\frac{1-B^2 M^2+\sqrt{B^4 M^4-14 B^2 M^2+1}}{2 B^2 M}\, .
\end{eqnarray}
By substituting $r=1/u$ and $r^\prime =-u^\prime /u^2$, Eq.~\eqref{radialPhi} can be rewritten  as follows
\begin{eqnarray}
    \left( \frac{du}{d\phi} \right)^2 = \frac{1}{b^2}-\mathcal{Q}u^2\equiv G(u)\, .\label{firstorbiteq}
\end{eqnarray}
The complete expression for $G(u)$ is
    \begin{eqnarray}
    G(u)=&&\frac{1}{b^2}-\mathcal{Q}u^2\nonumber\\=&&\frac{1}{b^2}-u^2+2 M u^3-B^2+B^4 M^2+2 B^2 M u\nonumber\\&&+B^2 M^2 u^2\, .\label{Frbitfunc}
\end{eqnarray}   

Using Eq.~\eqref{firstorbiteq}, the azimuthal deflection angle $\varphi$ of photons around the SBR BH can be expressed as~\cite{Peng_2021,Sharipov:2026mxc}
\begin{eqnarray}
\varphi(b) = 
\begin{cases} 
\displaystyle \ \ \
\int\limits_0^{1/r_h} \frac{du}{\sqrt{G(u)}}\, ,  \quad & b < b_c\  \\[12pt]
\displaystyle \ \ \
\quad\infty\ ,  \quad & b = b_c\  \\[12pt]
\displaystyle
2 \int\limits_0^{1/r_{\min}} \frac{du}{\sqrt{G(u)}}\ , \quad & b > b_c\ .
\end{cases}\label{deflec}
\end{eqnarray}
Here, $r_h$ and $r_{min}$ represent the event horizon radius and the turning point radius, respectively. $r_{min}$ is determined by solving the equation $G(1/r)=0$.

To obtain ray-tracing plots around the SBR BH, we use Eqs.~\eqref{firstorbiteq} and~\eqref{Frbitfunc} to determine the orbital equation and the initial conditions $(u_0$ and $u_0^\prime)$ required to solve it
\begin{eqnarray}
   \frac{d^2u}{d\phi^2}+u&=& B^2 M^2 u+B^2 M+3 M u^2\,\label{orbit} ,\\
   u_0&=&\lim_{r \to \infty}u=\lim_{r \to \infty}\frac{1}{r}=0\, ,\\
   u_0^\prime&=&\lim_{u \to 0} \left| \frac{du}{d\phi} \right| =\lim_{u \to 0} \sqrt{G(u)}\nonumber\\&=&\sqrt{\frac{1}{b^2}-B^2+B^4 M^2}\, .
\end{eqnarray}

Here, we determine the initial conditions by assuming that the photons come from an infinite distance $(r \to \infty)$. These photons arrive in parallel and are deflected around the BH due to its strong gravitational field. According to Eq.~\eqref{deflec}, the deflection depends on the impact parameter $b$. Therefore, the photon trajectories around the SBR BH, shown in Fig.~\ref{fig:ray1}, are highlighted in different colors corresponding to different ranges of the impact parameter $b$. This figure is generated from the orbital equation given in Eq.~\eqref{orbit}. In Fig.~\ref{fig:ray1}, the ray tracing is plotted at different values of magnetic field strength $B=0, 0.03,$ and $0.05$, and we can observe how increasing the magnetic field strength affects the photon trajectories. The trajectory of a ray corresponding to the critical impact parameter $b_c$ is represented by the black dashed curve. The figure depicts the event horizon region as a black disk, with the photon sphere indicated by a black dashed circle. The upper-right side of the plots illustrates a bundle of parallel rays coming from infinity. As $B$ increases, this bundle expands because the spacetime becomes Melvin-like at infinity (when $B=0$, it reduces to Minkowski spacetime), and the initial condition required to solve the orbital equation is modified: $u_0^\prime=\sqrt{1/b^2-B^2+B^4 M^2}$ (Schwarzschild BH case: $u_0^\prime=1/b$).

Based on the number of ray crossings $N$ with the vertical axis of the BH depicted in Fig.~\ref{fig:ray1}, the rays can be classified into three classes, namely: direct emission ($N=1$, deflection angle $\pi/2\le\varphi<3\pi/2$), lensed emission ($N=2, 3\pi/2\le\varphi<5\pi/2$), and photon-ring ($N \ge 3, \varphi\geq5\pi/2$); see, e.g., Refs.~\cite{Zhang:2024hix,Peng_2021,Gralla_2019,XamidovAccretion2025}. In Table~\ref{tab:nb}, the boundaries of this classification for the SBR BH at different values of the magnetic field strength $B$ can be listed in terms of the impact parameter $b$. 
 
In Table~\ref{tab:nb}, we also provide the values of the significant quantities: event horizon $r_h$, photon sphere $r_{ph}$, ISCO radius $r_{ISCO}$, and $b_c$ for different values of magnetic field strength $B$. Almost all of these quantities increase with increasing magnetic field strength $B$, but the value of the critical impact parameter $b_c$ decreases. Increasing values of $r_h, r_{ph}$ and $r_{ISCO}$ indicate an increase in gravitational influence. Since the spacetime at infinity is not Minkowski, the parallel bundle of photons expands (see Fig.~\ref{fig:ray1}). Consequently, as the value of $B$ increases, the critical impact parameter $b_c$ and the ranges of trajectories of individually classified photons shift toward smaller values. That is, for $B=0.05$, $b_c=5.157$ and the lensed emissions lie in $b\in(4.958,5.131)\cup(5.173,6.112)$, Whereas for $B=0$ (the Schwarzschild BH case), $b_c=5.196$ and the lensed emissions lie in $b\in(5.015,5.188)\cup(5.228,6.168)$.
\begin{figure*} 
    \centering
    \includegraphics[scale=0.55]{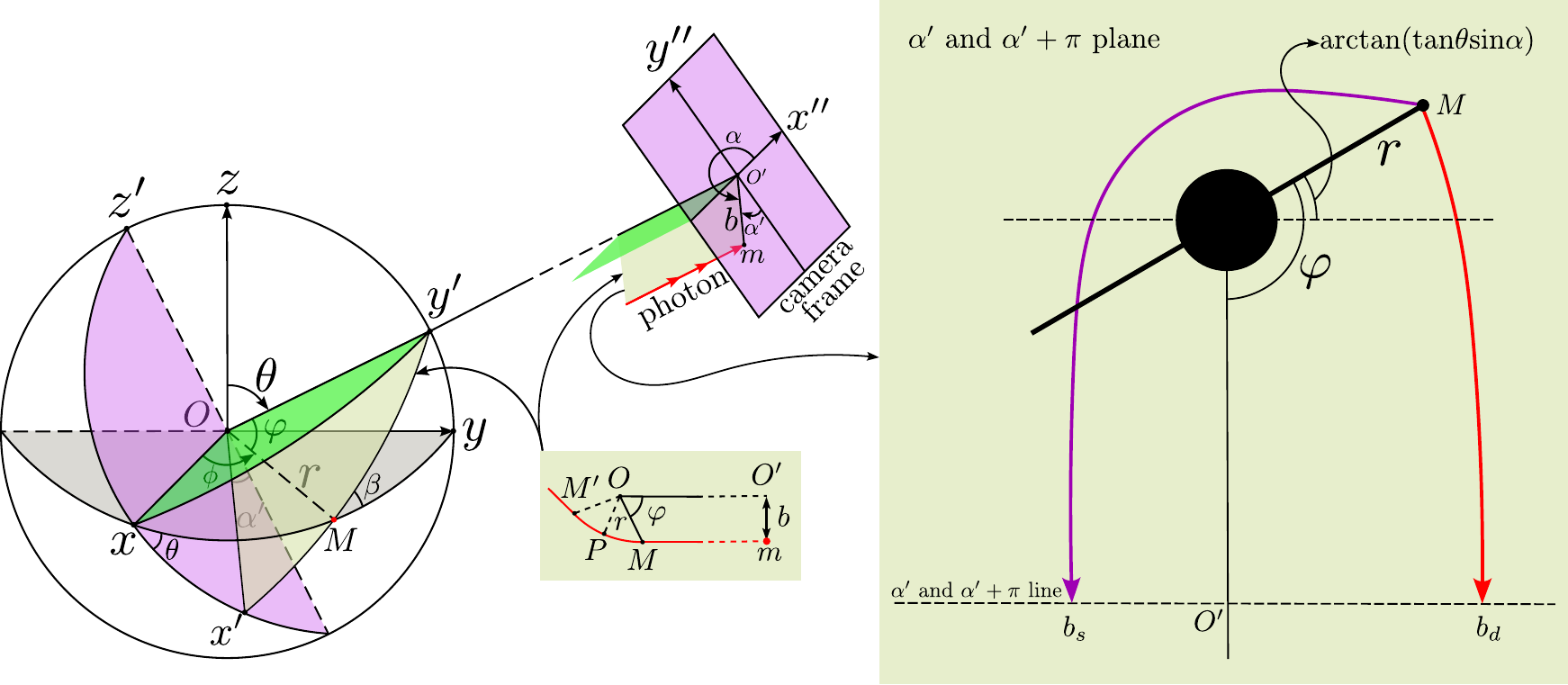}
    \caption{Schematic illustration of the coordinate system employed to construct the accretion disk image \cite{Luminet1979,XamidovAccretion2025}.
    }
 \label{fig:coordinate}
\end{figure*}
  
\begin{figure*}[!htb]
    \centering
    \includegraphics[width=0.48\linewidth]{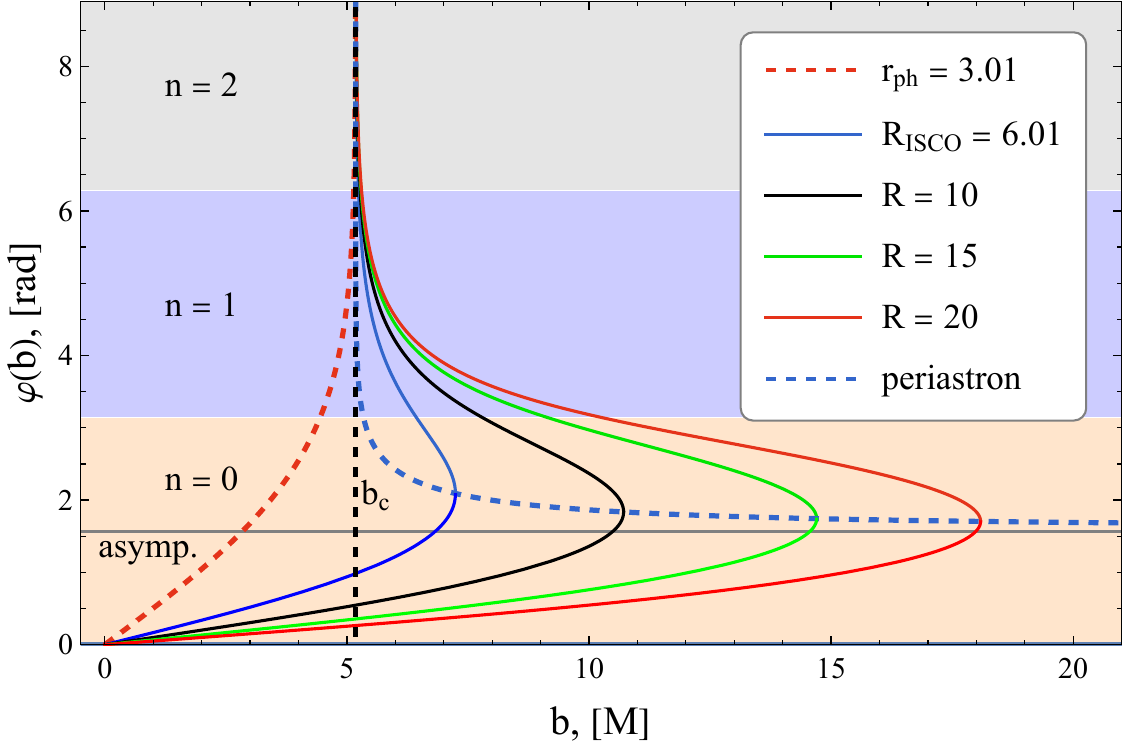}\hspace{0.5cm}
    \includegraphics[width=0.48\linewidth]{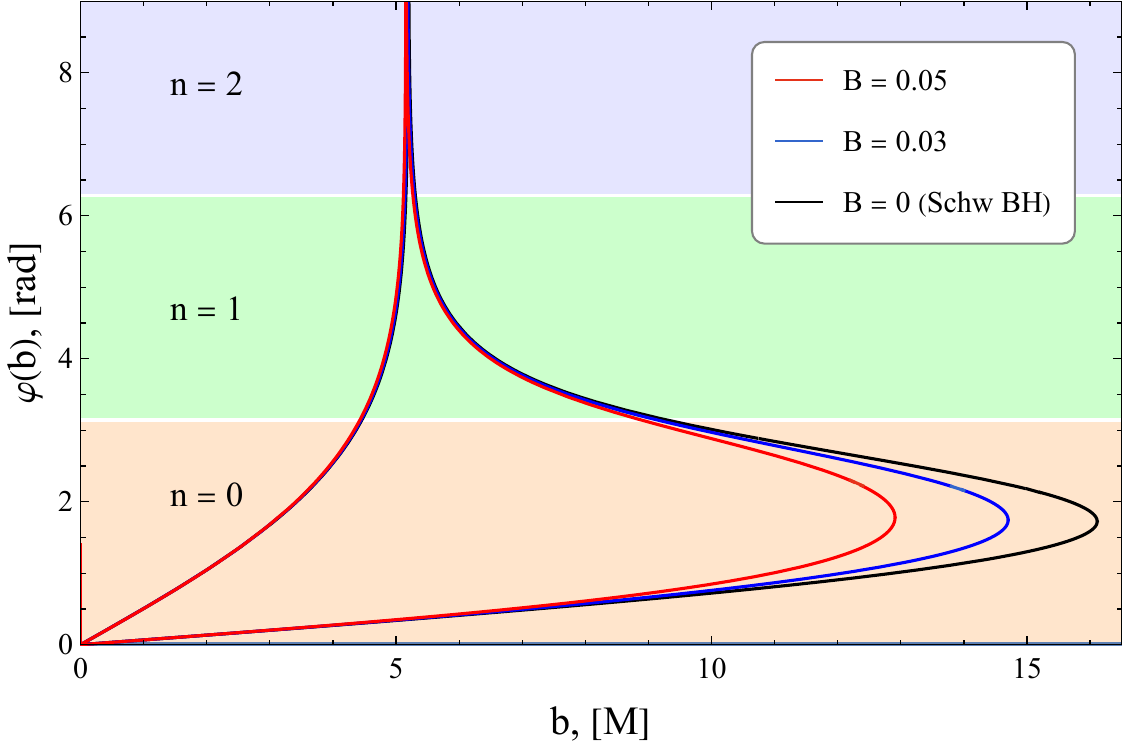}
    \caption{The plots illustrate image formation diagrams for different values of the disk radii $R$ at fixed $B=0.03$ (left panel) and for different values of the magnetic field $B$ at fixed radius $R=15$ (right panel).}
    \label{fig:FBline1}
\end{figure*}

\section{IMAGES AND PHYSICAL PROPERTIES OF THIN ACCRETION DISK }\label{Sec:III}
\subsection{Image Formation of Accretion Disks}\label{subsec:A}

This section is devoted to the study of the radiation behavior of an accretion disk orbiting the SBR BH. The disk is described within the Novikov–Thorne model~\cite{ Cui24,1973blho.conf..343N} as an optically thick and geometrically thin configuration composed of orbiting high-temperature plasma, gas, and dust. Under this assumption, the vertical thickness $h$ is much smaller than the radial distance $r$ ($h\ll r$), so the vertical structure can be ignored in comparison with the radial extent of the disk. Observationally, accretion disks are mainly characterized by their electromagnetic emission, radial temperature profile, and luminosity. In this work, particular attention is paid to evaluating how the magnetic field strength $B$ influences the disk’s apparent image, energy flux, and the associated gravitational redshift effects.

The formation of images for a thin accretion disk is examined within an observational coordinate system, illustrated in Fig.~\ref{fig:coordinate}. In this setup, the observer is located at an infinite distance in a spherical coordinate system $(r, \theta, \phi)$ centered on the BH, where the origin $(r=0)$ corresponds to the central singularity. In the corresponding camera frame $O'x''y''$, a photon emitted from an image-plane point $m(b, \alpha)$ propagates perpendicularly toward the BH and intersects the disk at the position $M(r, \pi/2, \phi)$. Utilizing the principle of optical path reversibility, the trajectory of a photon emitted from the disk at $M(r, \pi/2, \phi)$ precisely retraces the path leading to the image point $m(b, \alpha)$ in the camera frame.

In this work, two distinct types of observed images are considered. The direct image, specified by the parameters $(b_d, \alpha)$, is produced by photons that propagate from the accretion disk to the observer’s detector without completing any orbital motion around the BH (see the right-hand panel of Fig.~\ref{fig:coordinate}), whereas the secondary image, labeled $(b_s, \alpha + \pi)$, arises from photons initially emitted in the direction opposite to the observer. These photons are strongly deflected by the BH’s gravitational field, curving around it before reaching the camera plane. A schematic illustration of these photon trajectories for both image types is presented in Fig.~\ref{fig:coordinate}. 

From an application of the spherical sine theorem to $\triangle Myy' $ and $ \triangle Mxx' $, we derive the following equations~\cite{XamidovAccretion2025}:
\begin{eqnarray}
    \frac{\sin\varphi}{\sin\frac{\pi}{2}} = \frac{\sin(\frac{\pi}{2}-\theta)}{\sin\beta}\, ,
\end{eqnarray}
\begin{eqnarray}
    \frac{\sin(\frac{\pi}{2}-\varphi)}{\sin\theta} = \frac{\sin(\frac{\pi}{2}-\alpha^\prime)}{\sin\beta}\, .
\end{eqnarray}

Given the condition that $\alpha+\alpha^\prime=3\pi/2$ and applying the relationships derived earlier, the azimuthal angle $\varphi$ can be written in the following form~\cite{you2024,Sharipov:2025yfw}:
\begin{eqnarray}
\varphi = \frac{\pi}{2} + \arctan(\tan\theta \sin\alpha).
\end{eqnarray}

The azimuthal angle $\varphi_n$, which characterizes the $n^{th}$ order image, is defined by the following equation~\cite{you2024,XamidovAccretion2025}:
\begin{widetext}
\begin{eqnarray} \label{Eq:nthimage}
\varphi_n = 
\begin{cases}
\frac{n}{2} 2\pi + (-1)^n \left[ \frac{\pi}{2} + \arctan(\tan\theta \sin\alpha) \right], & \text{for $n$ even}, \\[6pt]
\frac{n+1}{2} 2\pi + (-1)^n \left[ \frac{\pi}{2} + \arctan(\tan\theta \sin\alpha) \right], & \text{for $n$ odd}\, .
\end{cases} 
\end{eqnarray}
\end{widetext}
Here, $n=0$ represents the direct image, while $n=1$ represents the secondary image of the accretion disk. We assume that a photon is emitted from the camera located at spatial infinity, characterized by the coordinates $(b, \alpha)$. As it propagates toward the BH and intersects a circular orbit of radius $r$ within the accretion disk, the total deflection angle along its trajectory can be calculated~\cite{you2024,XamidovAccretion2025}
\begin{figure*}
\begin{tabular}{ccc}
  \includegraphics[scale=0.32]{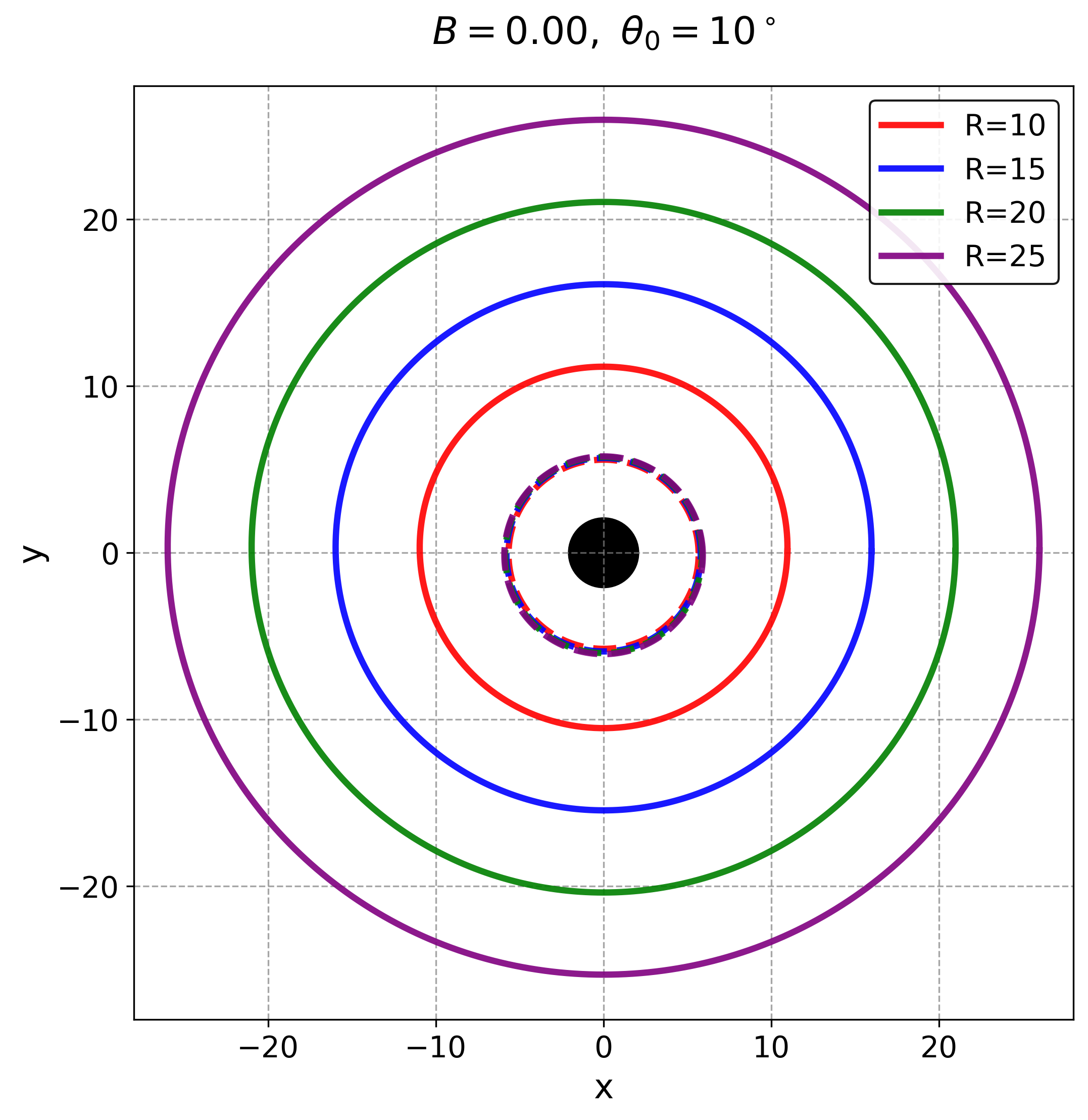}\hspace{-0.1cm}
  \includegraphics[scale=0.32]{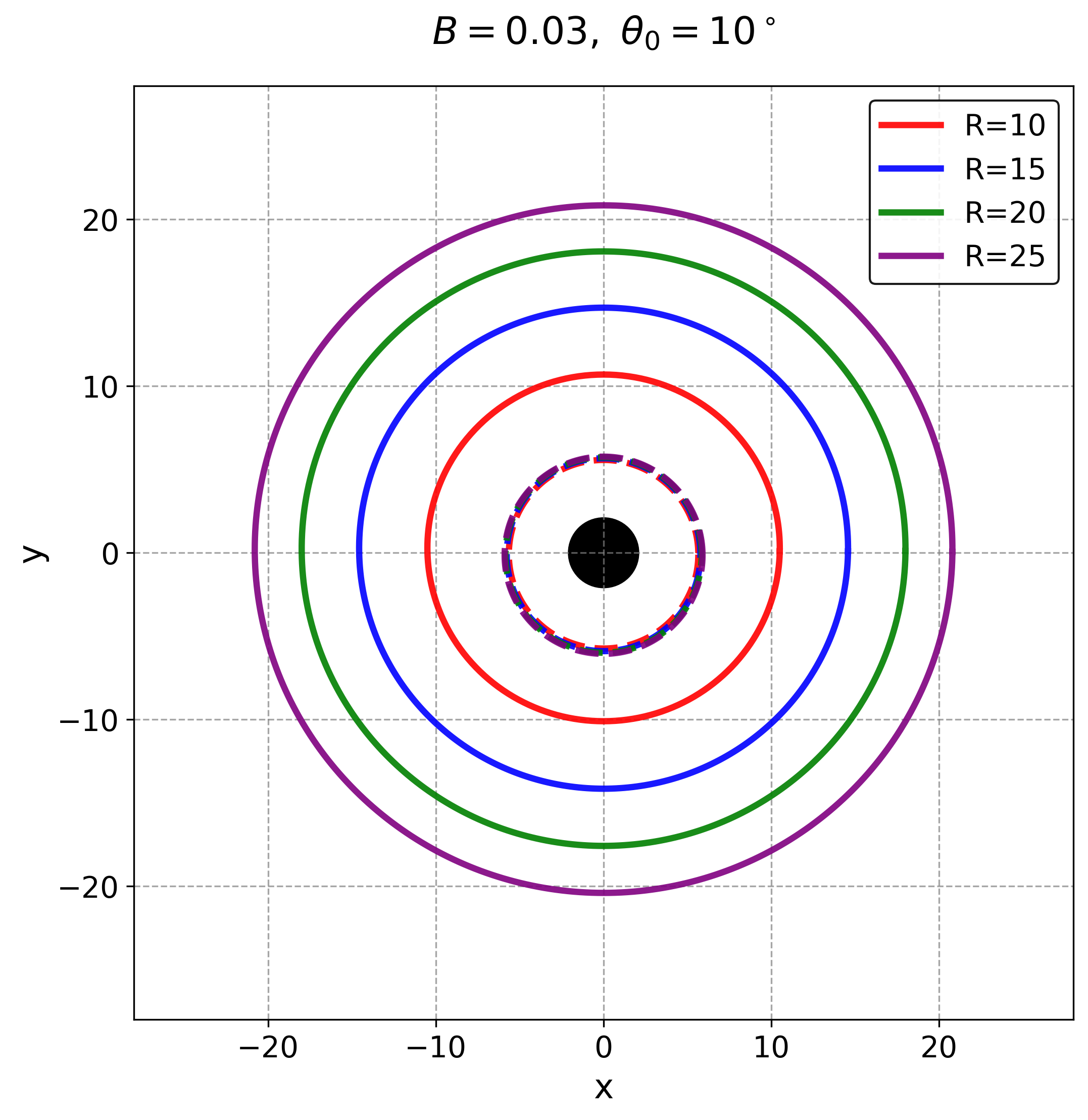}\hspace{-0.1cm}
  \includegraphics[scale=0.32]{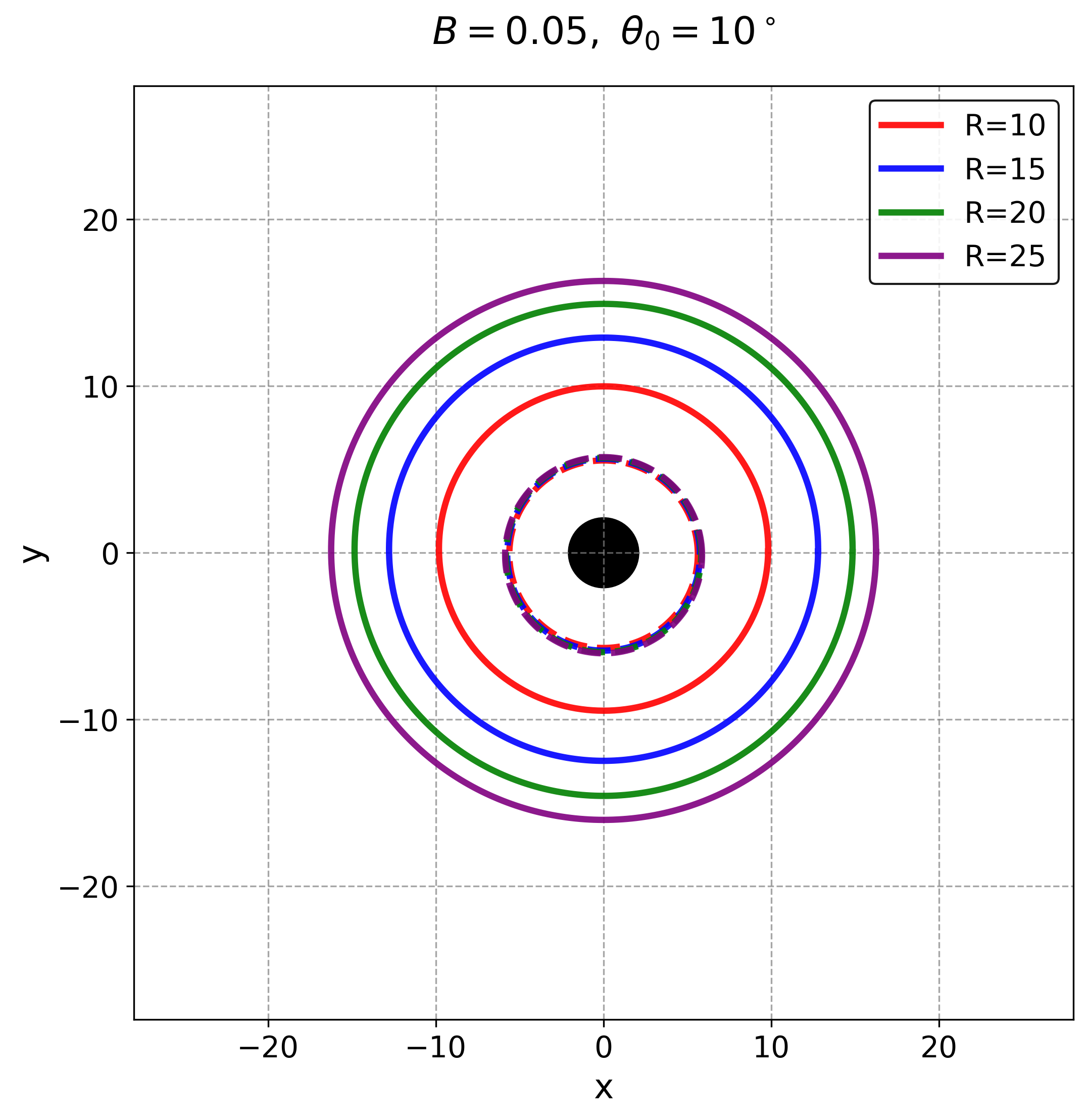}\\
  \includegraphics[scale=0.32]{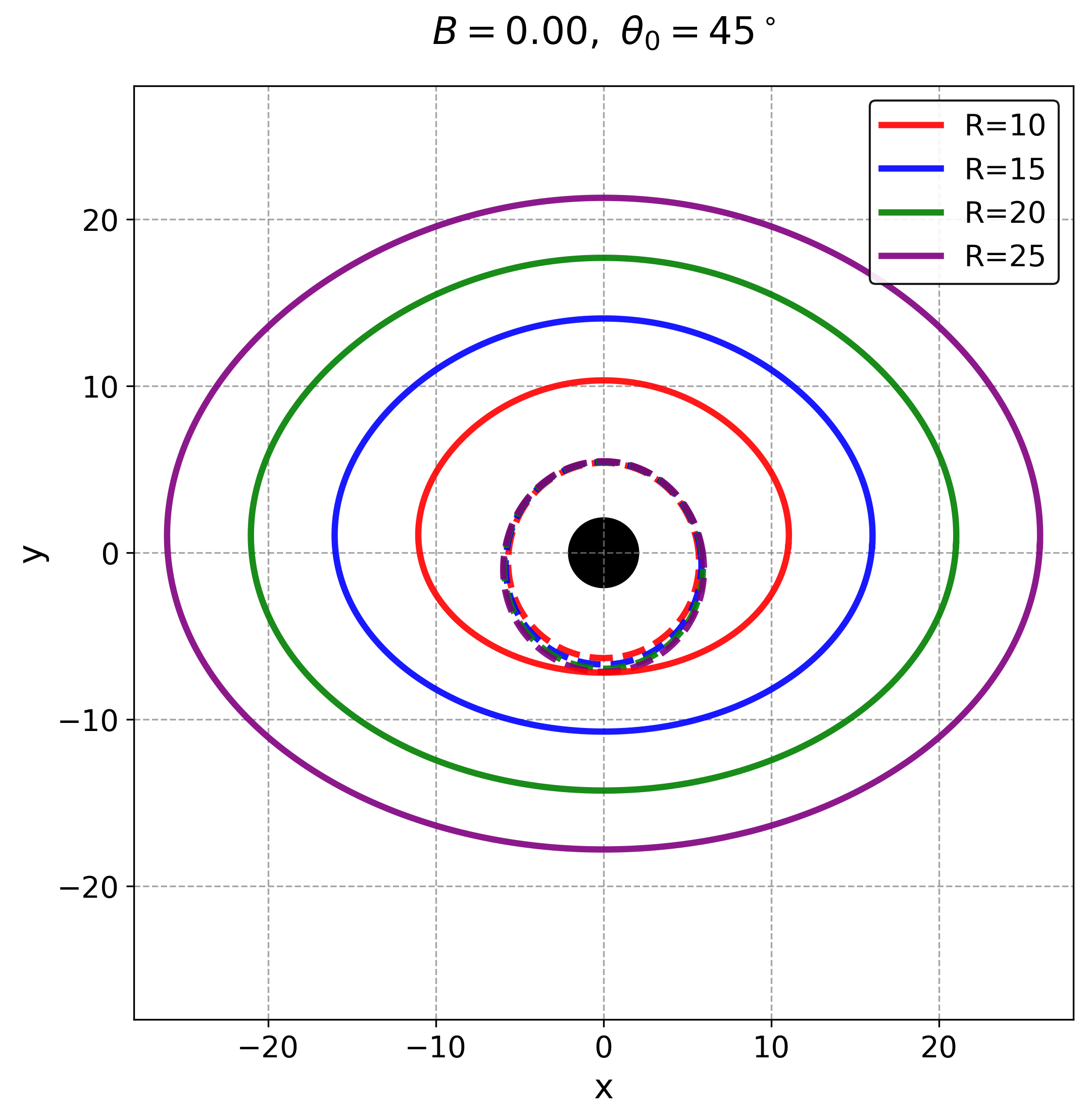}\hspace{-0.1cm}
  \includegraphics[scale=0.32]{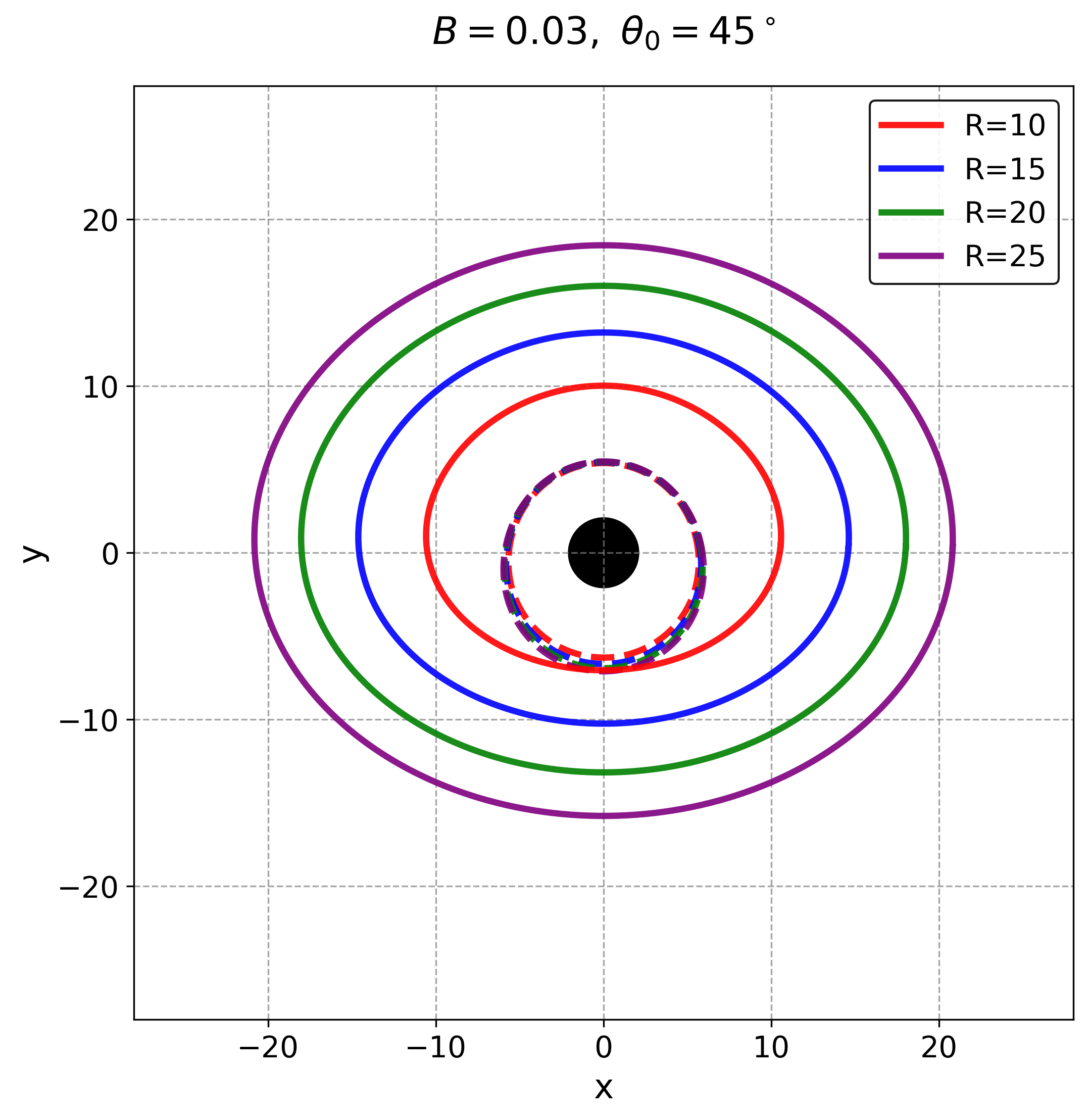}\hspace{-0.1cm}
  \includegraphics[scale=0.32]{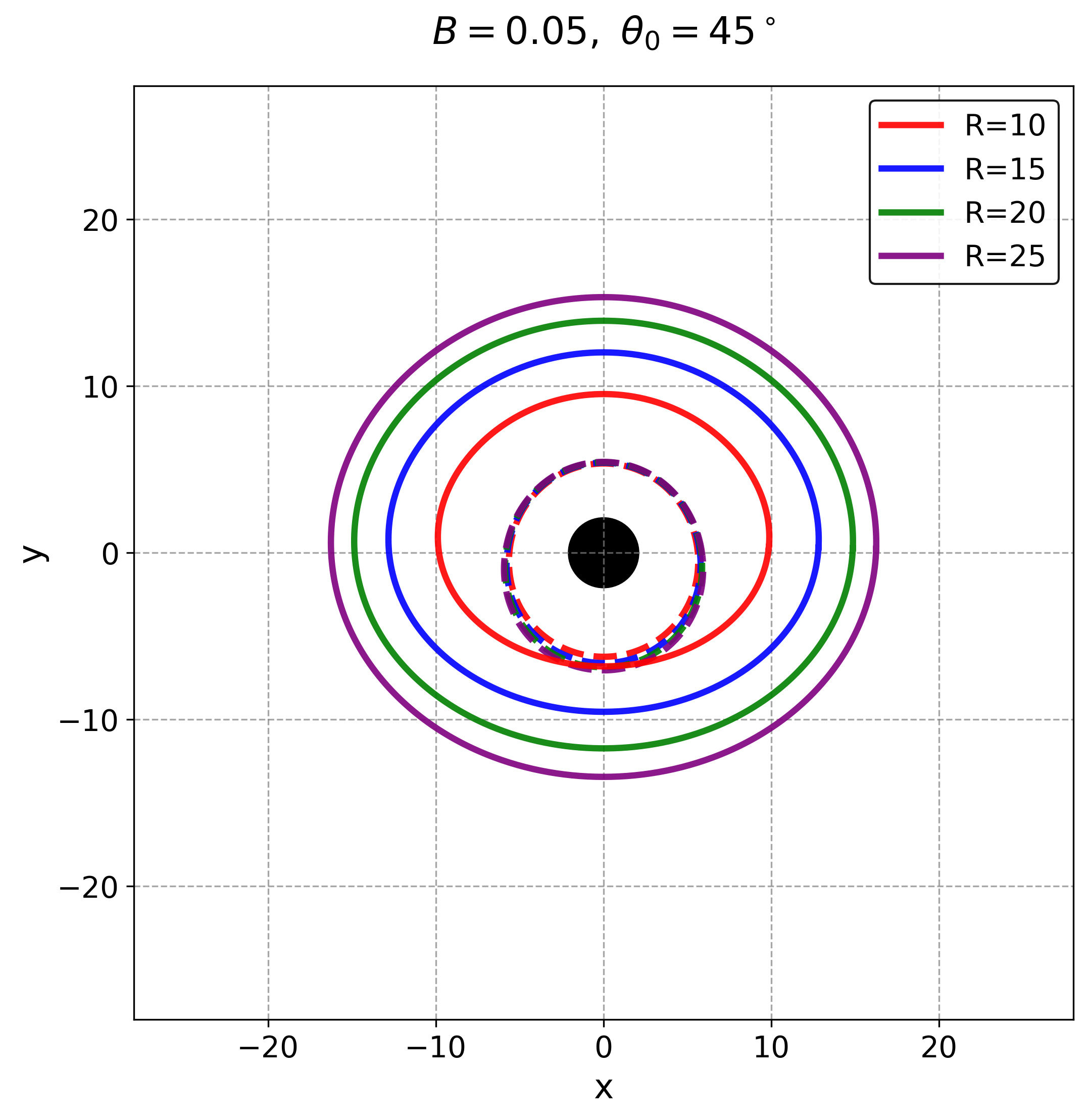}\\
  \includegraphics[scale=0.32]{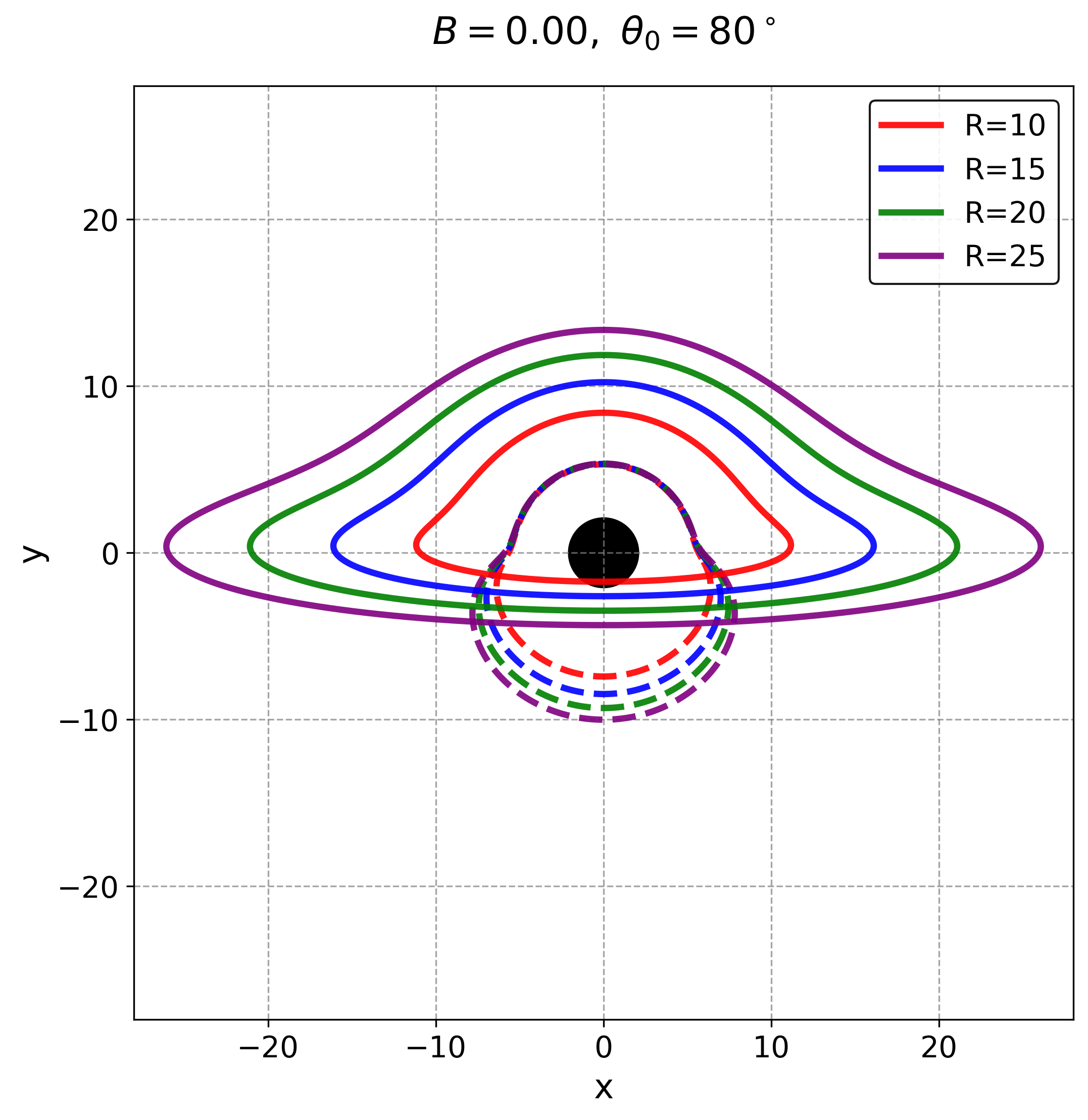}\hspace{-0.1cm}
  \includegraphics[scale=0.32]{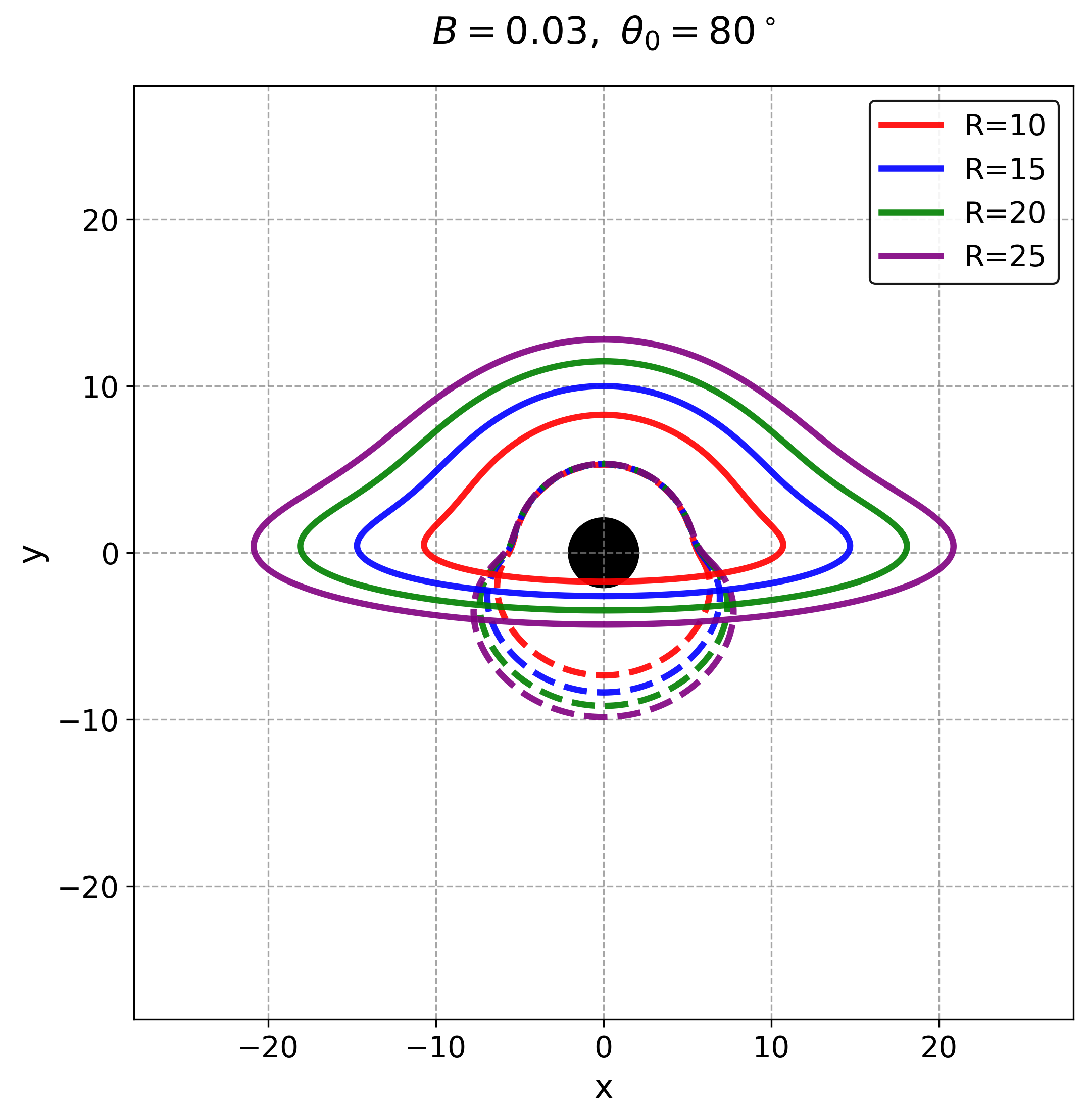}\hspace{-0.1cm}
  \includegraphics[scale=0.32]{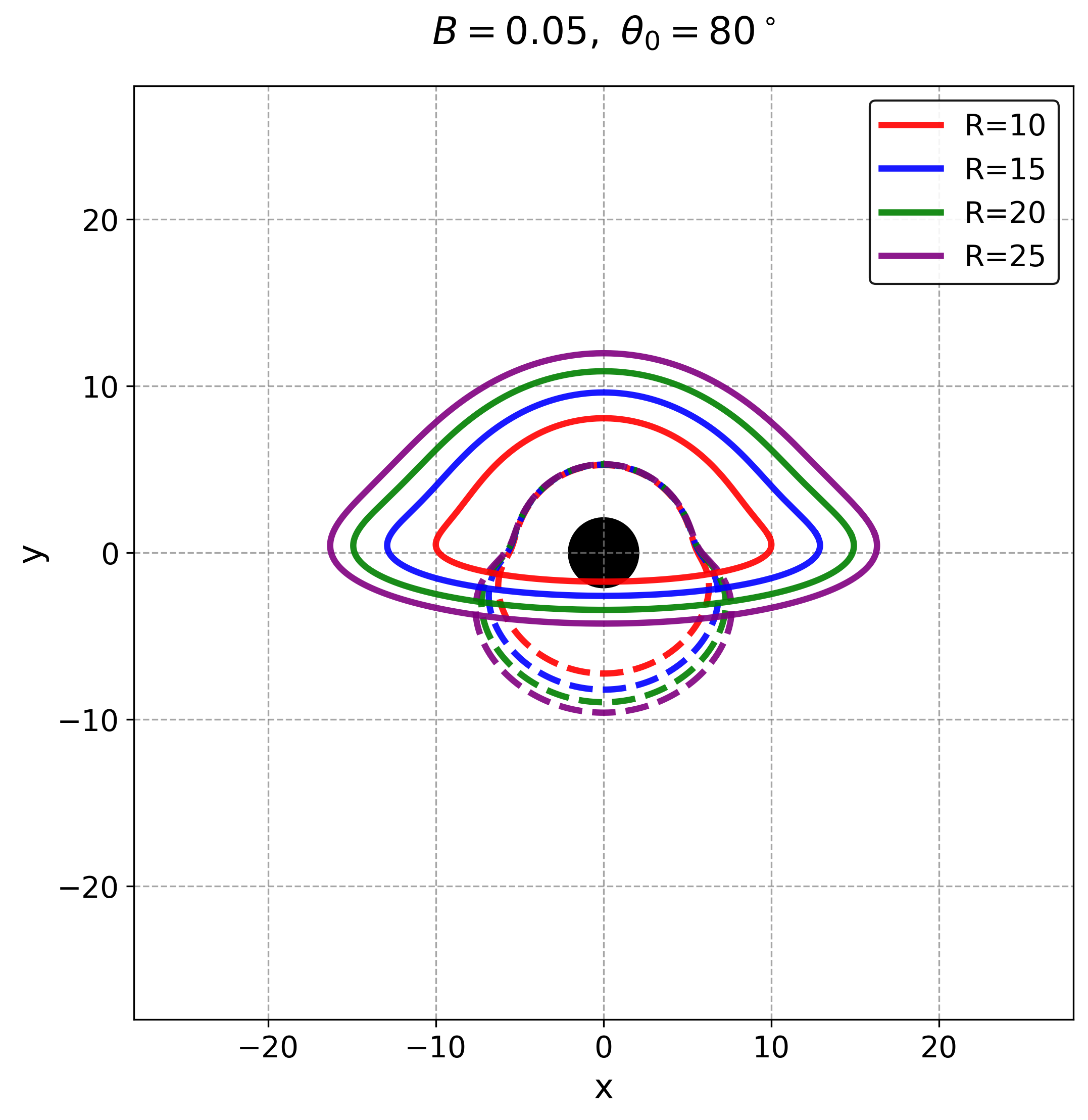}
  \end{tabular}
	\caption{\label{fig:accrthin} Direct (solid curves) and secondary (dashed curves) images of the accretion disk around the SBR BH for distinct disk radii $R$. The plots in each column correspond to different values of the magnetic field $B$, while rows correspond to a fixed inclination angle $\theta_0$.
   }
\end{figure*}
\begin{eqnarray} \label{Eq:phi1disk}
\varphi_1(b) = \int^{1/r}_{0} \frac{1}{\sqrt{G(u)}} du, 
\end{eqnarray}
\begin{eqnarray}\label{Eq:phi2disk}
\varphi_2(b) = 2 \int^{1/r_{\min}}_{0} \frac{1}{\sqrt{G(u)}} du - \int^{1/r}_{0} \frac{1}{\sqrt{G(u)}} du.
\end{eqnarray}
\begin{figure*}[!htb]
   \centering
  \includegraphics[width=0.48\linewidth]{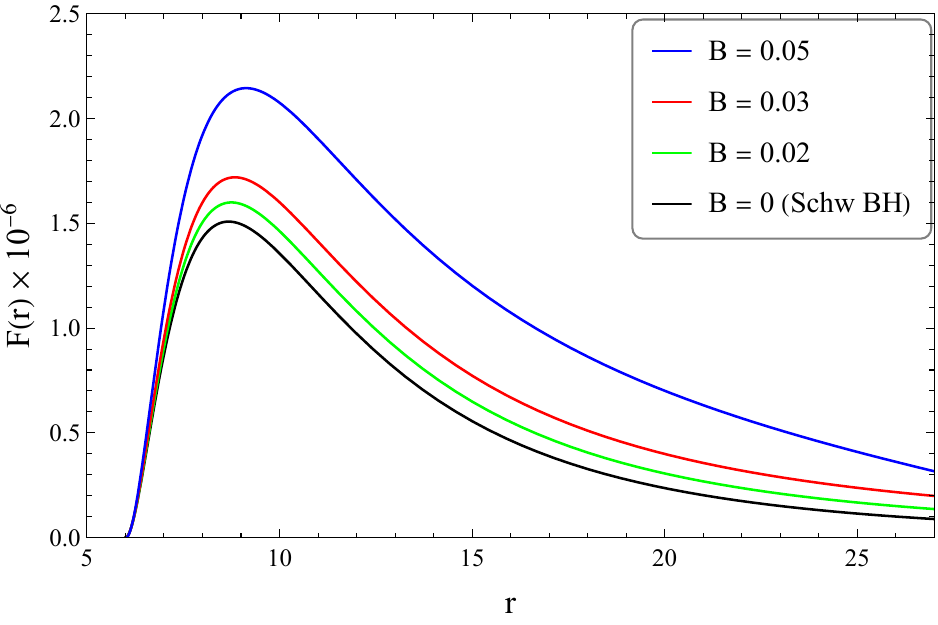}\hspace{0.5cm}
 \includegraphics[width=0.48\linewidth]{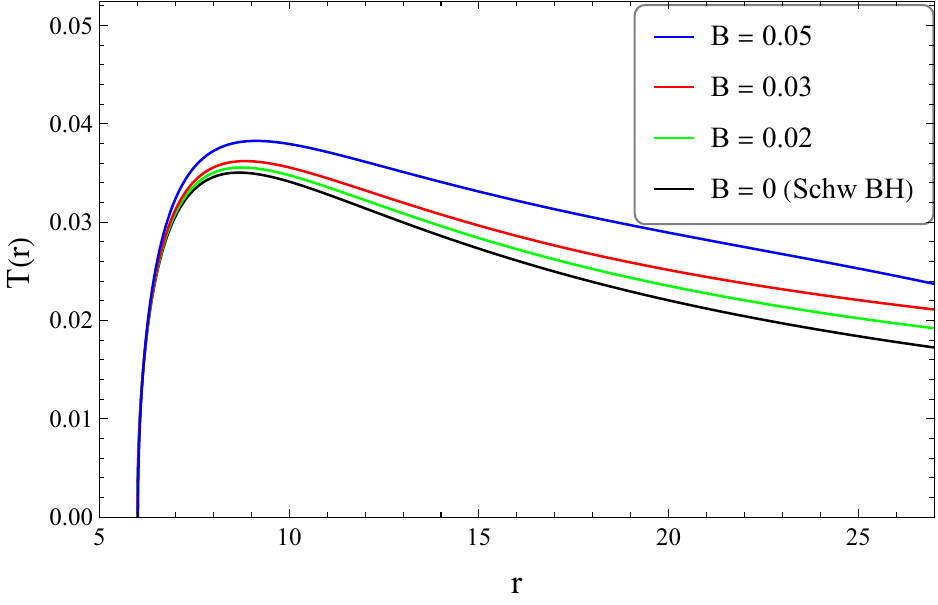}
\caption{The radial distributions of energy flux $F(r)$ (left) and temperature 
$T(r)$ (right) on the accretion disk for different values of magnetic field $B$.}
\label{fig:Flux}
\end{figure*}

In Fig.~\ref{fig:FBline1}, we present image formation diagrams obtained using the deflection angle given by Eqs. $\varphi_1(b)$ and $\varphi_2(b)$, where $b$ represents the impact parameter. Each colored section represents a specific image order $n$. The left panel displays the deflection angle for photons emitted from the accretion disk radii $(R=6.01, 10, 15, 20)$, with the magnetic field fixed at $B=0.03$. The corresponding maximum impact parameters are $b_{max}=7.25$, $10.71$, $14.7$, and $18.09$, respectively. By analyzing the curves obtained for these different radii, we can see that the images of order $n=0$ and $n=1$ are distinct, whereas higher-order images $(n\geq2)$ overlap. The blue dashed curve, calculated from $\varphi_3$ and approaching the asymptotic line $\varphi=\pi/2$, represents the periastron and intersects the points where the impact parameter b reaches its maximum on the solid curves~\cite{you2024} 
\begin{eqnarray}
\varphi_3(b) = \int^{1/r_{\min}}_{0} \frac{1}{\sqrt{G(u)}} du\, . 
\end{eqnarray}
In the right panel, image formation diagrams are illustrated for different values of the magnetic field $B$, with $R=15$ fixed. The case $B=0$ corresponds to the Schwarzschild BH. As can be clearly seen from the figure, the maximum impact parameter $b_{\max}$ decreases as $B$ increases, indicating that the magnetic field $B$ enhances the central gravitational influence. This variation affects only the direct image (orange region); The change is almost imperceptible in the higher-order images (green and gray regions).

Solving the Eqs.~\eqref{Eq:nthimage}-\eqref{Eq:phi2disk} numerically allows us to determine the apparent location of the accretion disk on the observer’s plane, characterized by the coordinates $(b, \alpha)$ for both direct and higher-order images~\cite{Ziqiang25,Gyulchev20}. Fig.~\ref{fig:accrthin}  illustrates the direct and secondary images of circular orbits with different radii around the SBR BH. As the magnetic field $B$ increases, the size of the direct image decreases, while the secondary image remains largely unchanged. This reduction in image size occurs because the photons forming the image from the orbit are deflected more (see Fig.~\ref{fig:FBline1} ).
\begin{figure}
    \centering
\includegraphics[width=1\linewidth]{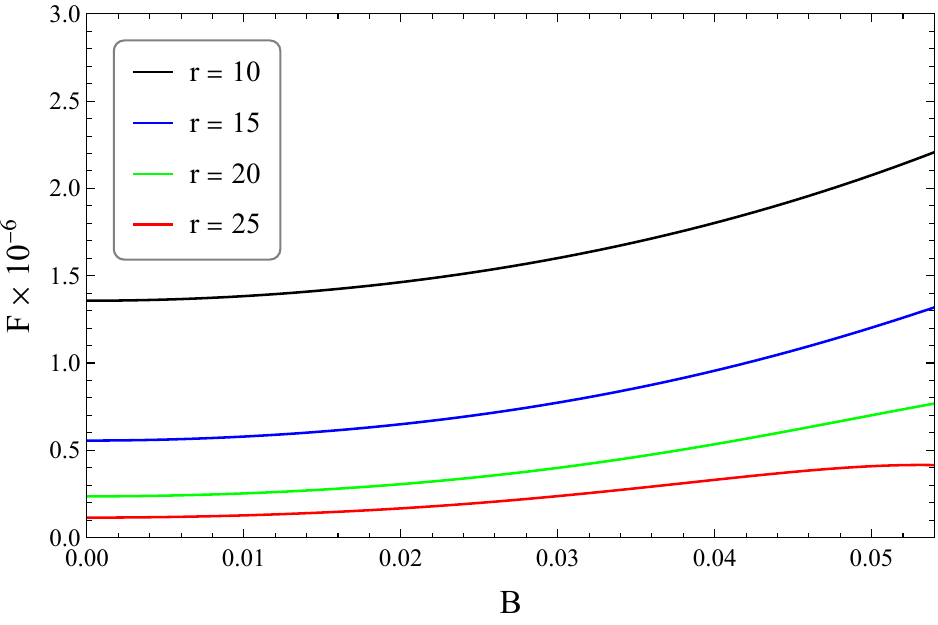}
    \caption{The plot demonstrates energy flux $F$ as a function of magnetic field strength $B$.}
    \label{fig:F_B}
\end{figure}

\subsection{Magnetized Keplerian Dynamics and Disk Efficiency}

The SBR BH metric presented in the \cref{Sec:II} provides a unique opportunity to study accretion disks in a background magnetic field strong enough to modify the spacetime geometry itself. Unlike perturbative approaches that treat the magnetic field as a test field on a fixed Schwarzschild background, the SBR metric incorporates the field non-perturbatively, allowing us to derive analytic expressions for the Keplerian flow, the innermost stable circular orbit (ISCO), and the radiative efficiency of a thin accretion disk.

We consider a geometrically thin, optically thick accretion disk in the equatorial plane ($\theta = \pi/2$). For a stationary, axisymmetric flow, the four-velocity of the fluid is $u^\mu = (u^t, 0, 0, u^\phi)$. In the equatorial plane, the metric functions simplify considerably. From Eq.~(1), $P(\theta=\pi/2)=1$, and $\Omega^2(\theta=\pi/2) = 1 + B^2 r^2$, so the metric \cref{eq:metric} becomes
\begin{equation}
ds^2 = \frac{1}{1+B^2 r^2} \left[ -\mathcal{Q} dt^2 + \frac{dr^2}{\mathcal{Q}} + r^2 d\phi^2 \right],
\end{equation}
where $\mathcal{Q} = (1 + B^2 r^2)\left(1-\frac{2M}{r}-B^2 M^2\right)$. Therefore, the relevant metric components are
\begin{eqnarray}
g_{tt} &=& -\frac{\mathcal{Q}}{1+B^2 r^2}, \quad g_{\phi\phi} = \frac{r^2}{1+B^2 r^2}, \nonumber\\ g_{rr} &=& \frac{1}{\mathcal{Q}(1+B^2 r^2)}.
\end{eqnarray}

The spacetime Killing vectors, $\partial_t$ and $\partial_\phi$, corresponding to conserved energy and angular momentum per unit rest mass: $E = -g_{tt} u^t$ and $L = g_{\phi\phi} u^\phi$. From these, we obtain $u^t = E(1+B^2 r^2)/\mathcal{Q}$ and $u^\phi = L(1+B^2 r^2)/r^2$. The four-velocity normalization condition $g_{\mu\nu}u^\mu u^\nu = -1$ yields the radial equation $\frac{1}{2}(u^r)^2 + V_{\text{eff}}(r) = 0$, with the effective potential
\begin{equation}
V_{\text{eff}}(r) = \frac{1}{2g_{rr}}\left(1 + \frac{L^2}{g_{\phi\phi}} + \frac{E^2}{g_{tt}}\right).
\end{equation}

For circular orbits, we require $u^r = 0$ and $\partial_r V_{\text{eff}} = 0$. These conditions give, respectively,
\begin{equation}
1 + \frac{L^2}{g_{\phi\phi}} + \frac{E^2}{g_{tt}} = 0, \quad \frac{d}{dr}\left(1 + \frac{L^2}{g_{\phi\phi}} + \frac{E^2}{g_{tt}}\right) = 0.
\end{equation}

The angular velocity $\Omega_K = u^\phi/u^t = d\phi/dt$ can be derived without explicitly solving for $E$ and $L$. From the definitions of the conserved quantities, $L = -E g_{\phi\phi} \Omega_K / g_{tt}$. Substituting this into the first circular orbit condition and simplifying leads to the general relation $\Omega_K^2 = -(dg_{tt}/dr)/(dg_{\phi\phi}/dr)$ for metrics with $g_{t\phi}=0$. Computing the derivatives for SBR BH metric components gives
\begin{align}
\frac{dg_{tt}}{dr} &= -\frac{\mathcal{Q}'(1+B^2 r^2) - 2B^2 r \mathcal{Q}}{(1+B^2 r^2)^2}, \\
\frac{dg_{\phi\phi}}{dr} &= \frac{2r}{(1+B^2 r^2)^2},
\end{align}
we obtain
\begin{equation}
\Omega_K^2 = \frac{\mathcal{Q}'(1+B^2 r^2) - 2B^2 r \mathcal{Q}}{2r}.
\end{equation}
Using $\mathcal{Q} = (1+B^2 r^2)(1-2M/r-B^2 M^2)$ and its radial derivative $r$ $\mathcal{Q}' = 2B^2 r (1-2M/r-B^2 M^2) + (1+B^2 r^2)(2M/r^2)$, the square of the angular velocity simplifies to
\begin{equation}
\Omega_K^2 = \frac{M}{r^3} \left(1+B^2 r^2\right)^2.
\end{equation}
Taking the positive square root for prograde orbits,
\begin{equation}
\Omega_K = \sqrt{\frac{M}{r^3}} \left(1+B^2 r^2\right). \label{eq:omega}
\end{equation}
For $B=0$, we recover the Schwarzschild Keplerian frequency $\Omega_K = \sqrt{M/r^3}$. The magnetic field introduces a multiplicative factor $(1+B^2 r^2)$, which increases the angular velocity compared to the unmagnetized case. This has profound implications for the disk dynamics.

The specific energy and angular momentum for circular orbits follow from the normalization condition. Using $u^t = 1/\sqrt{-(g_{tt} + g_{\phi\phi}\Omega_K^2)}$ and $E = -g_{tt}u^t$, we find
\begin{equation}
E = \frac{1-\frac{2M}{r}-B^2 M^2}{\sqrt{1-\frac{3M}{r} - B^2 M^2 - M B^2 r}}. \label{eq:energy}
\end{equation}
Similarly, from $L = g_{\phi\phi}\Omega_K u^t$,
\begin{equation}
L = \frac{\sqrt{Mr}}{\sqrt{1-\frac{3M}{r} - B^2 M^2 - M B^2 r}}. \label{eq:angmom}
\end{equation}
For $B=0$, these reduce to the standard Schwarzschild expressions $E = (1-2M/r)/\sqrt{1-3M/r}$ and $L = \sqrt{Mr}/\sqrt{1-3M/r}$.

The innermost stable circular orbit is determined by the condition $dL/dr = 0$, marking the transition from stable to unstable circular orbits. From Eq.~(\ref{eq:angmom}), it is convenient to work with $L^2 = Mr/D$, where
\begin{equation}
D = 1 - \frac{3M}{r} - B^2 M^2 - M B^2 r\, .
\end{equation}
The logarithmic derivative condition $d\ln L^2/dr = 0$ gives
\begin{equation}
\frac{1}{r} = \frac{1}{D}\left(\frac{3M}{r^2} - M B^2\right)\, ,
\end{equation}
which is further simplified as
\begin{equation}
1 - \frac{6M}{r} - B^2 M^2 = 0.
\end{equation}
Notice that this simplification is significant: it tells us that the magnetic field enters only through the constant $B^2 M^2$ term. Now,
Solving this for $r$ is now straightforward and gives
\begin{equation}
r = \frac{6M}{1 - B^2 M^2}\, .
\end{equation}

To see the leading effect of the magnetic field, we introduce the dimensionless parameter $\beta = BM$. For weak fields ($\beta \ll 1$), we expand:
\begin{equation}
r = 6M \left(1 - \beta^2\right)^{-1} = 6M \left(1 + \beta^2 + \beta^4 + \cdots\right).
\end{equation}
Thus, to leading order in $\beta^2$,
\begin{equation}\label{eq:isco}
r_{\text{ISCO}} = 6M \left(1 + \beta^2 + \mathcal{O}(\beta^4)\right).
\end{equation}
This result tells us that the ISCO moves outward as the magnetic field increases (also see \cref{tab:nb}). Intuitively, this makes sense: magnetic pressure pushes matter outward, requiring stable circular orbits to occur at larger radii. The effect is small: for a field with $\beta = 0.1$, the ISCO shifts outward by about $1\%$, yet it is a clean, analytic prediction of this exact magnetized spacetime. Unlike perturbative treatments that treat the field as a small correction on a fixed background, this result captures the full backreaction of the magnetic field on the geometry through the exact metric.

The radiative efficiency of the accretion disk, defined as the binding energy per unit rest mass at the ISCO, is $\eta = 1 - E(r_{\text{ISCO}})$. Using the energy expression from Eq.~(\ref{eq:energy}) and the ISCO radius from Eq.~(\ref{eq:isco}), we expand to order $\beta^2$.

Recall the energy expression from \cref{eq:energy} and substituting $r = 6M(1 + \beta^2 + \mathcal{O}(\beta^4))$ into this expression, we first compute the necessary expansions. Let $x = r/M = 6(1 + \beta^2 + \mathcal{O}(\beta^4))$. Then:
\begin{align}
\frac{2M}{r} &= \frac{2}{x} = \frac{2}{6(1 + \beta^2)} = \frac{1}{3}(1 - \beta^2 + \mathcal{O}(\beta^4)), \\
\frac{3M}{r} &= \frac{3}{x} = \frac{3}{6(1 + \beta^2)} = \frac{1}{2}(1 - \beta^2 + \mathcal{O}(\beta^4)), \\
M B^2 r &= \beta^2 x = 6\beta^2(1 + \beta^2) = 6\beta^2 + \mathcal{O}(\beta^4)\, .
\end{align}

Now the numerator of \cref{eq:energy} reads
\begin{align}
1 - \frac{2M}{r} - B^2 M^2 &= 1 - \frac{1}{3}(1 - \beta^2) - \beta^2 + \mathcal{O}(\beta^4)\, ,\nonumber\\ 
&= \frac{2}{3}\left(1 - \beta^2\right) + \mathcal{O}(\beta^4).
\end{align}
Next, evaluate the denominator inside the square root:
\begin{align}
1 - \frac{3M}{r} - B^2 M^2 - M B^2 r 
&= \frac{1}{2} - \frac{13}{2}\beta^2 + \mathcal{O}(\beta^4).
\end{align}
Therefore,
\begin{equation}
\sqrt{1 - \frac{3M}{r} - B^2 M^2 - M B^2 r} 
= \frac{1}{\sqrt{2}}\left(1 - \frac{13}{2}\beta^2\right) + \mathcal{O}(\beta^4).
\end{equation}

Now combine the numerator and denominator:
\begin{align}
E(r_{\text{ISCO}}) &= \frac{\frac{2}{3}\left(1 - \beta^2\right)}{\frac{1}{\sqrt{2}}\left(1 - \frac{13}{2}\beta^2\right)},\nonumber\\
&= \frac{2\sqrt{2}}{3}\left(1 - \beta^2\right)\left(1 + \frac{13}{2}\beta^2\right) + \mathcal{O}(\beta^4).
\end{align}
Expanding the product:
\begin{align}
E(r_{\text{ISCO}}) &= \frac{2\sqrt{2}}{3}\left(1 - \beta^2 + \frac{13}{2}\beta^2 + \mathcal{O}(\beta^4)\right),\nonumber\\ 
&= \frac{2\sqrt{2}}{3}\left(1 + \frac{11}{2}\beta^2 + \mathcal{O}(\beta^4)\right).
\end{align}

Thus, the radiative efficiency \cite{1973blho.conf..343N,1974ApJ...191..499P,1974ApJ...191..507T} becomes
\begin{align}
\eta &= 1 - E(r_{\text{ISCO}}),\nonumber\\  
&= 1 - \frac{2\sqrt{2}}{3}\left(1 + \frac{11}{2}B^2 M^2 + \mathcal{O}(B^4 M^4)\right). \label{eq:efficiency}
\end{align}

For $B=0$, $\eta \approx 0.057$, the familiar Schwarzschild BH value \cite{1973blho.conf..343N,Noble:2011wa}. When the magnetic field is turned on, the efficiency decreases, and this decrease becomes more pronounced as the field strength increases. For a field strong enough to modify the geometry ($\beta = BM \sim 0.1$), the efficiency drops to approximately $0.0053$, corresponding to a decrease of about $0.052$ or roughly $91\%$ relative to the unmagnetized case. This significant reduction arises because the outward shift of the ISCO increases the binding energy substantially, meaning considerably less energy is radiated away as the matter plunges into the BH.

While this dramatic decrease may seem surprising, it is a direct consequence of the exact magnetized spacetime geometry. The magnetic field, through its non-perturbative coupling to gravity, fundamentally alters the structure of circular orbits, pushing the ISCO outward to larger radii where the gravitational potential well is shallower. This prediction represents a clean, analytic result of this exact solution. The significant deviation in efficiency could, in principle, be detectable in the continuum spectra of highly magnetized active galactic nuclei, offering a potential observational signature that distinguishes this exact magnetized spacetime from both perturbative treatments and standard Schwarzschild accretion models. Future GRMHD simulations incorporating such strong magnetic fields may test this prediction, providing a unique discriminant for the non-perturbative coupling between magnetic fields and spacetime curvature.
\begin{figure*}
\begin{tabular}{ccc}
 \includegraphics[scale=0.34]{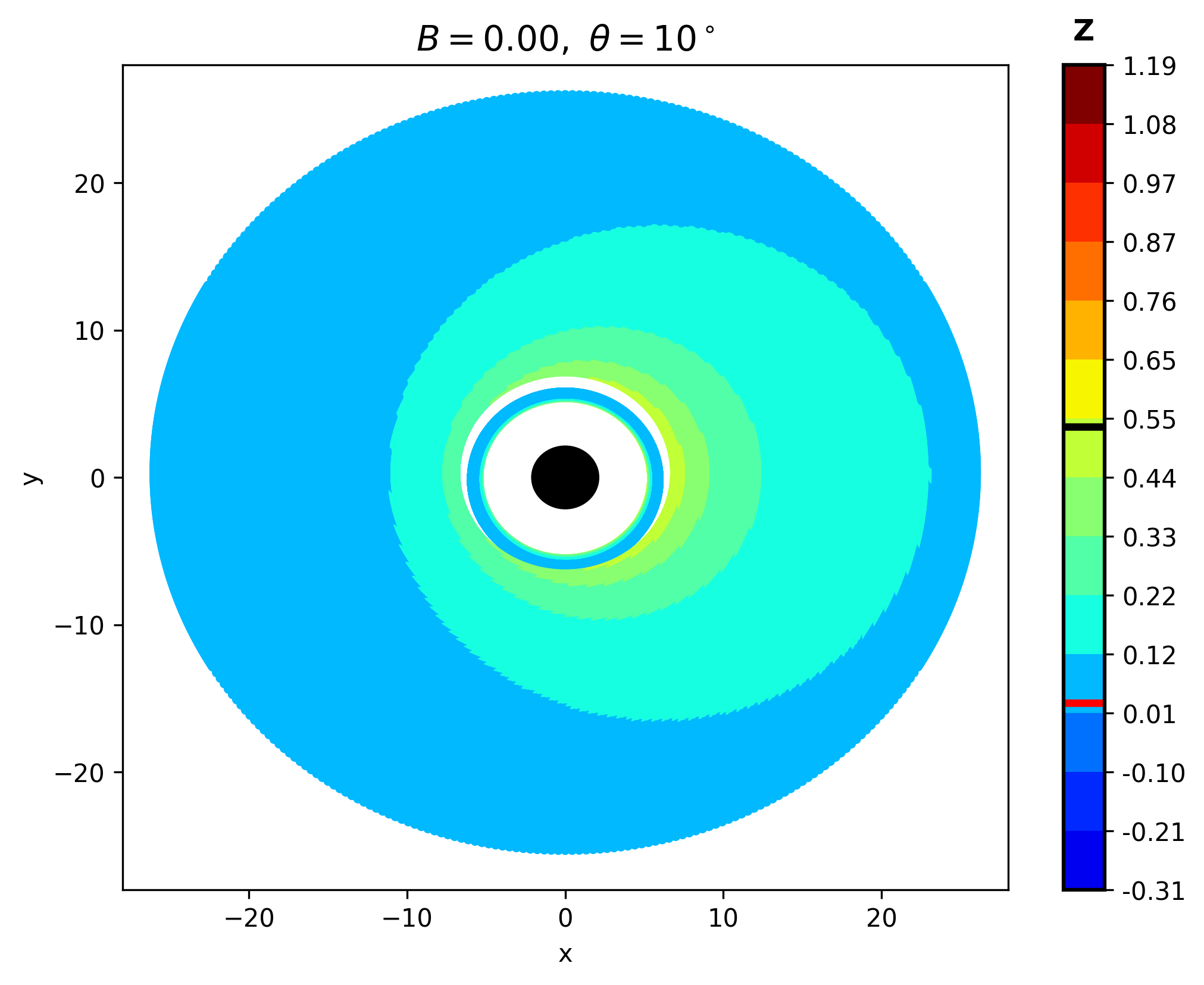}\hspace{-0.2cm}
 \includegraphics[scale=0.34]{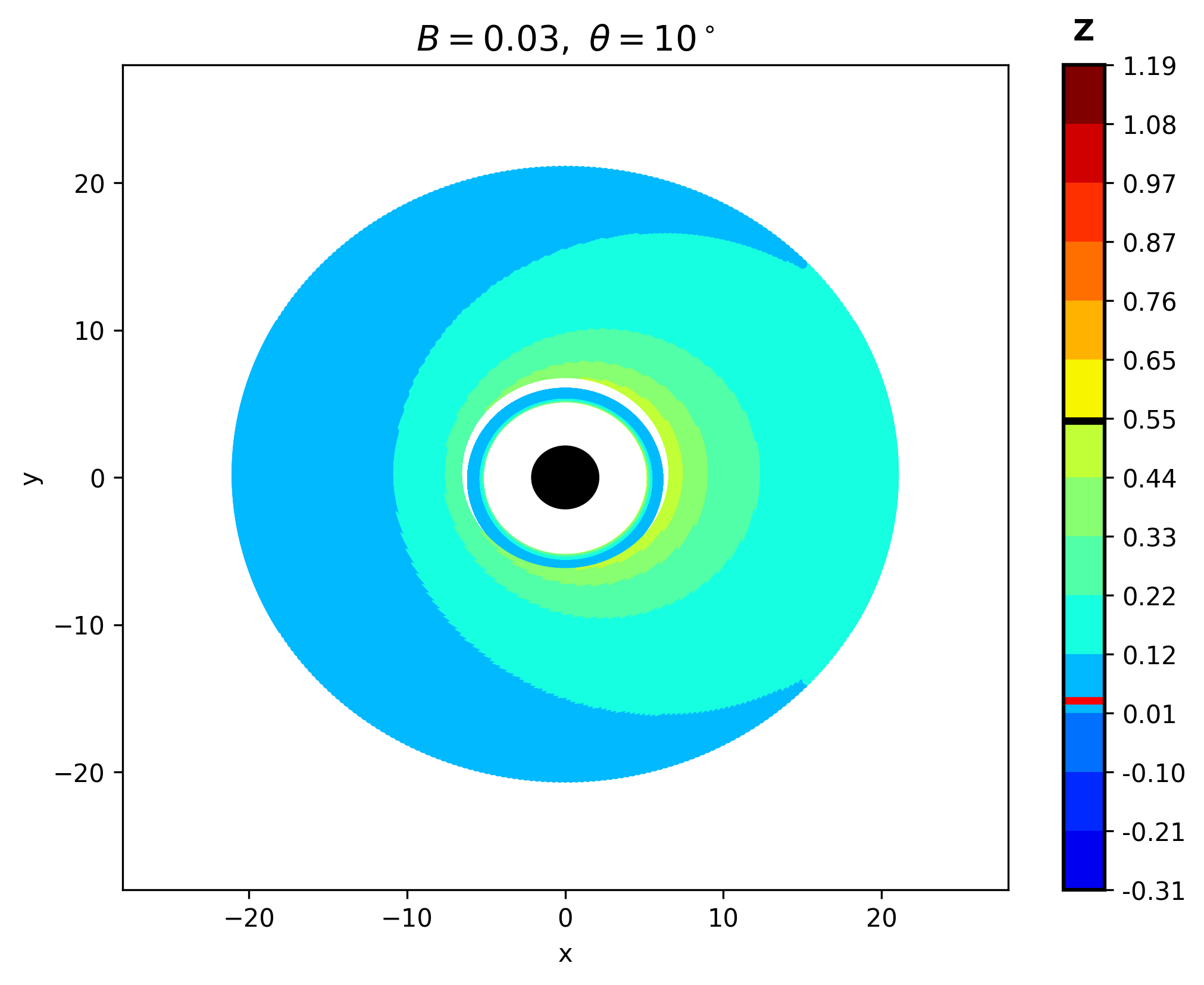}\hspace{-0.2cm}
 \includegraphics[scale=0.34]{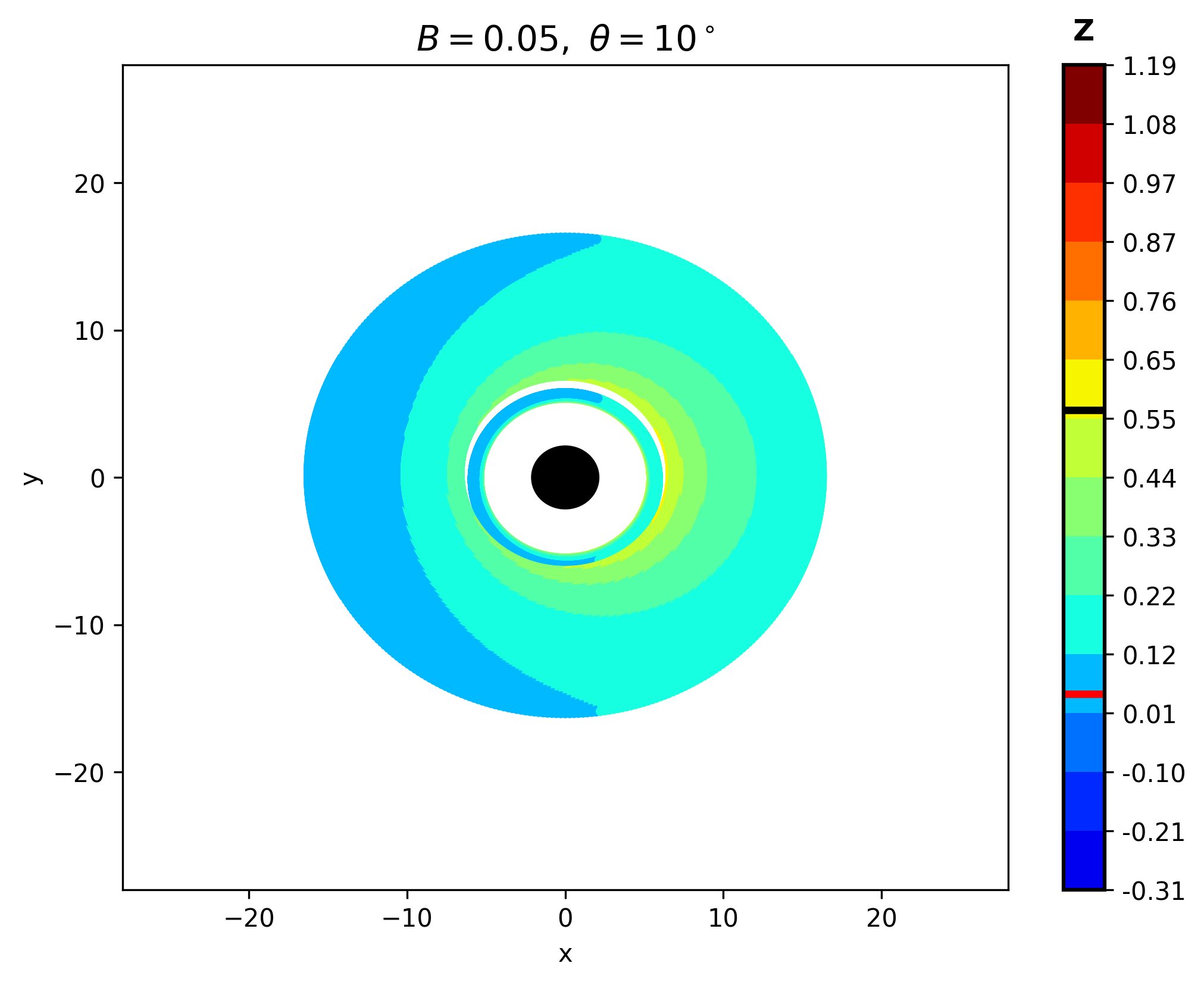}\\
 \includegraphics[scale=0.34]{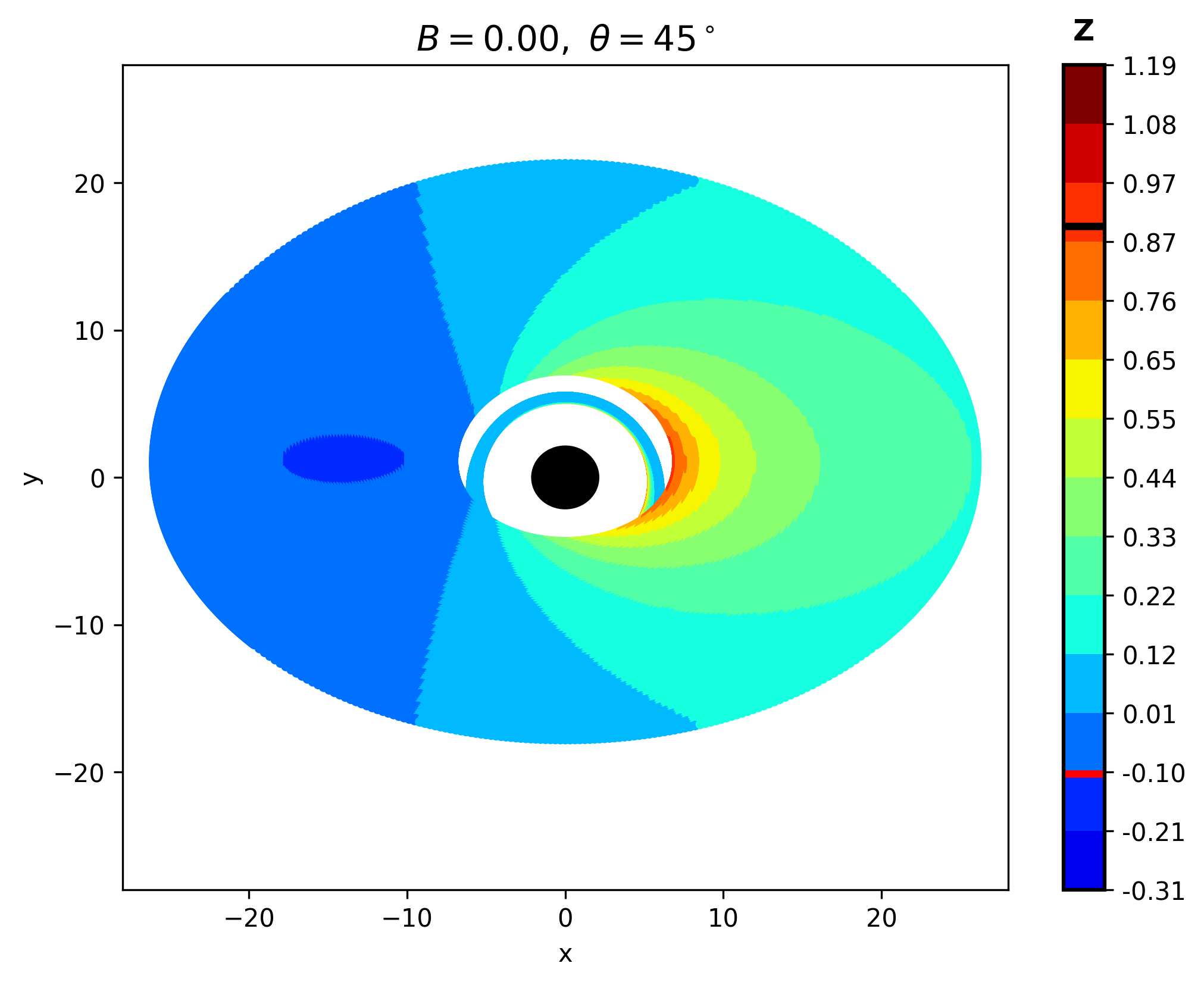}\hspace{-0.2cm}
 \includegraphics[scale=0.34]{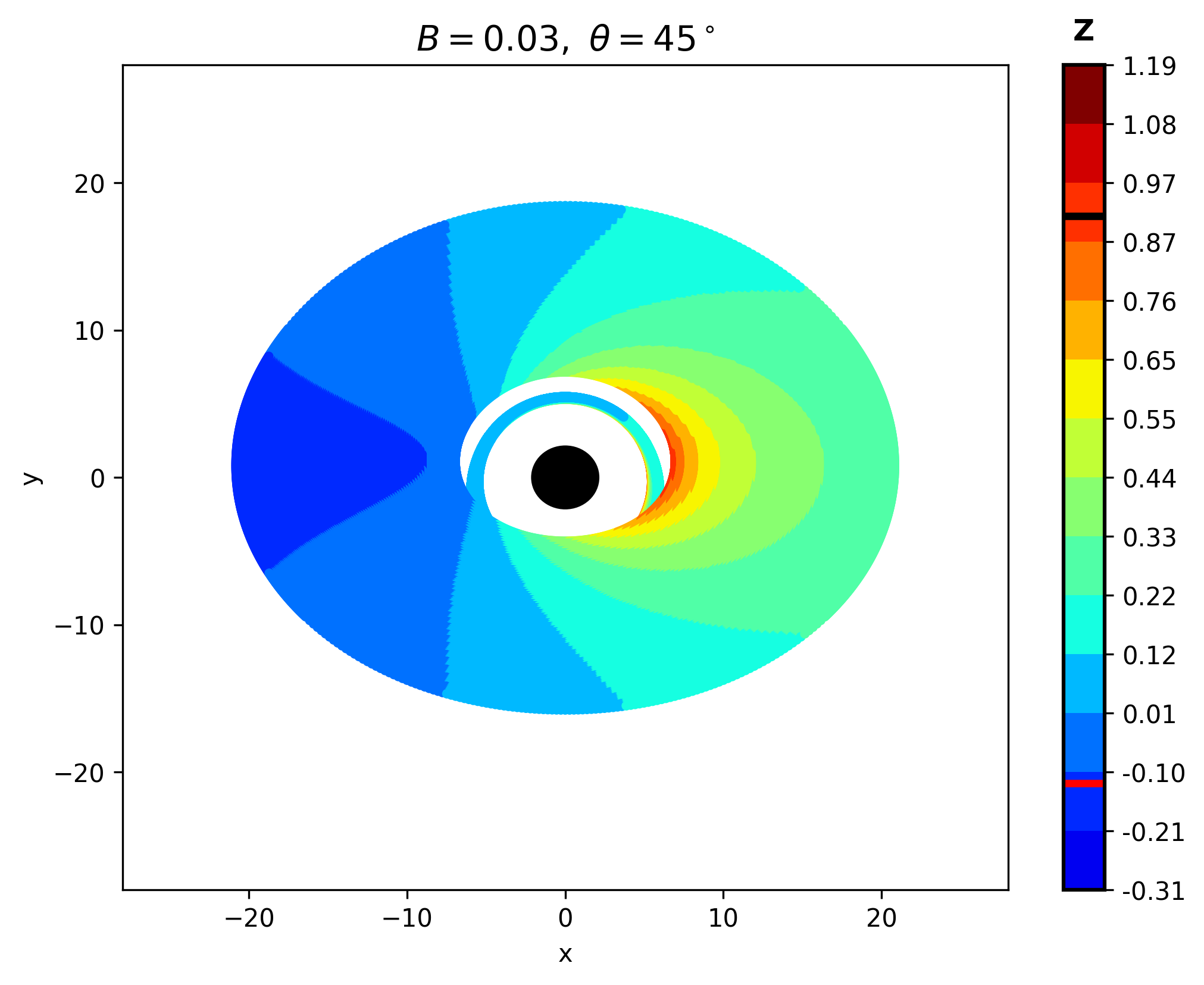}\hspace{-0.2cm}
 \includegraphics[scale=0.34]{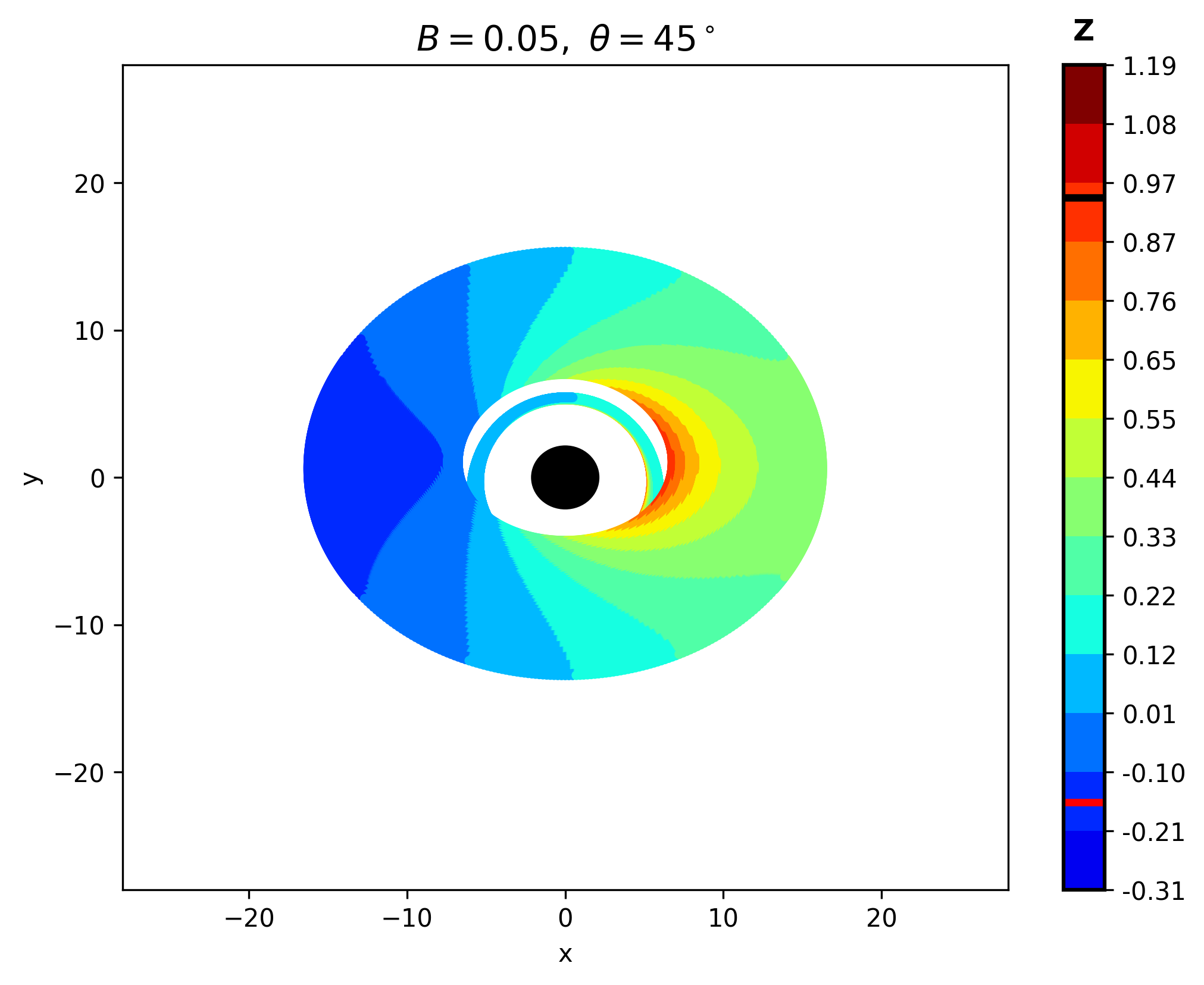}\\
 \includegraphics[scale=0.34]{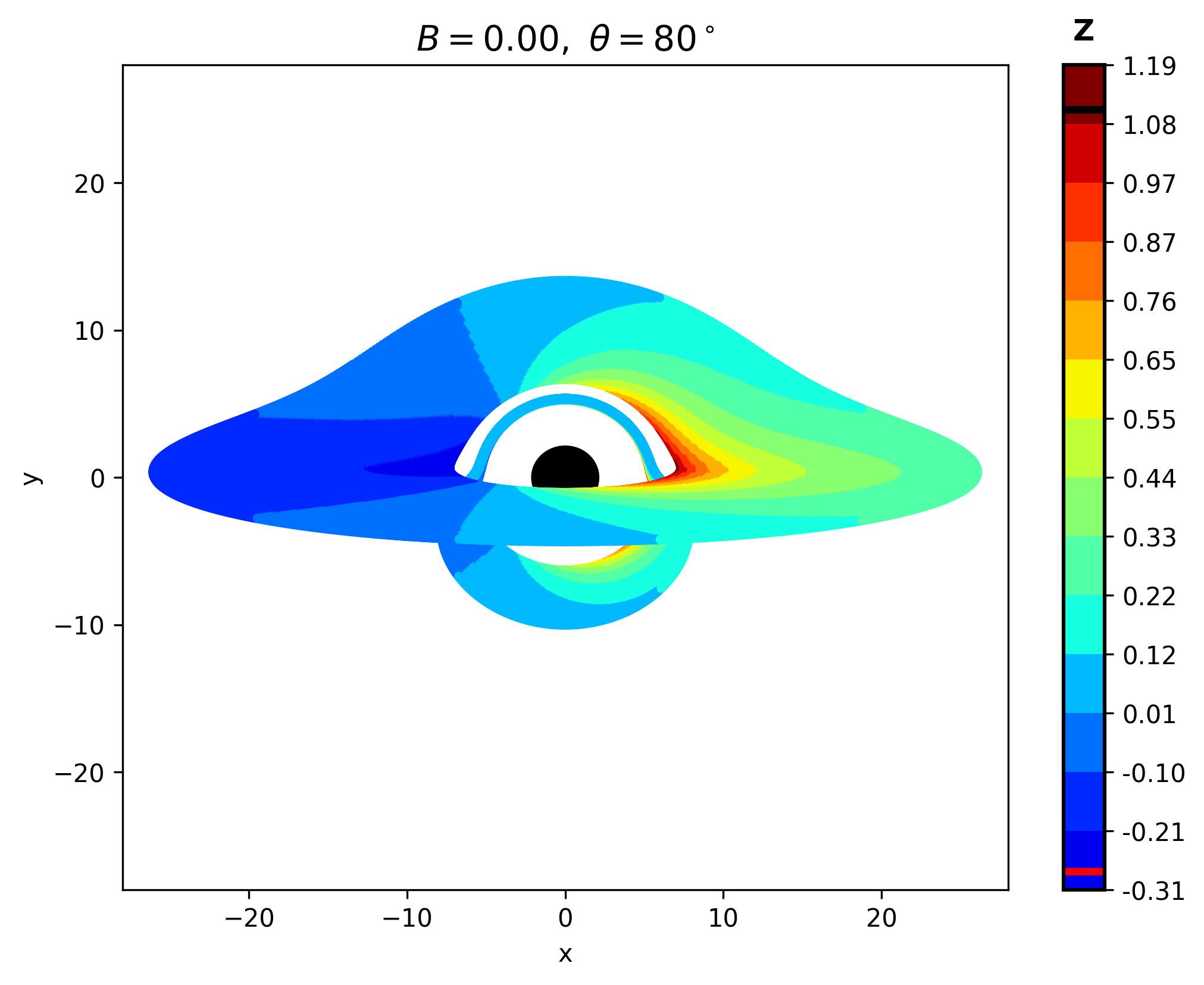}\hspace{-0.2cm}
 \includegraphics[scale=0.34]{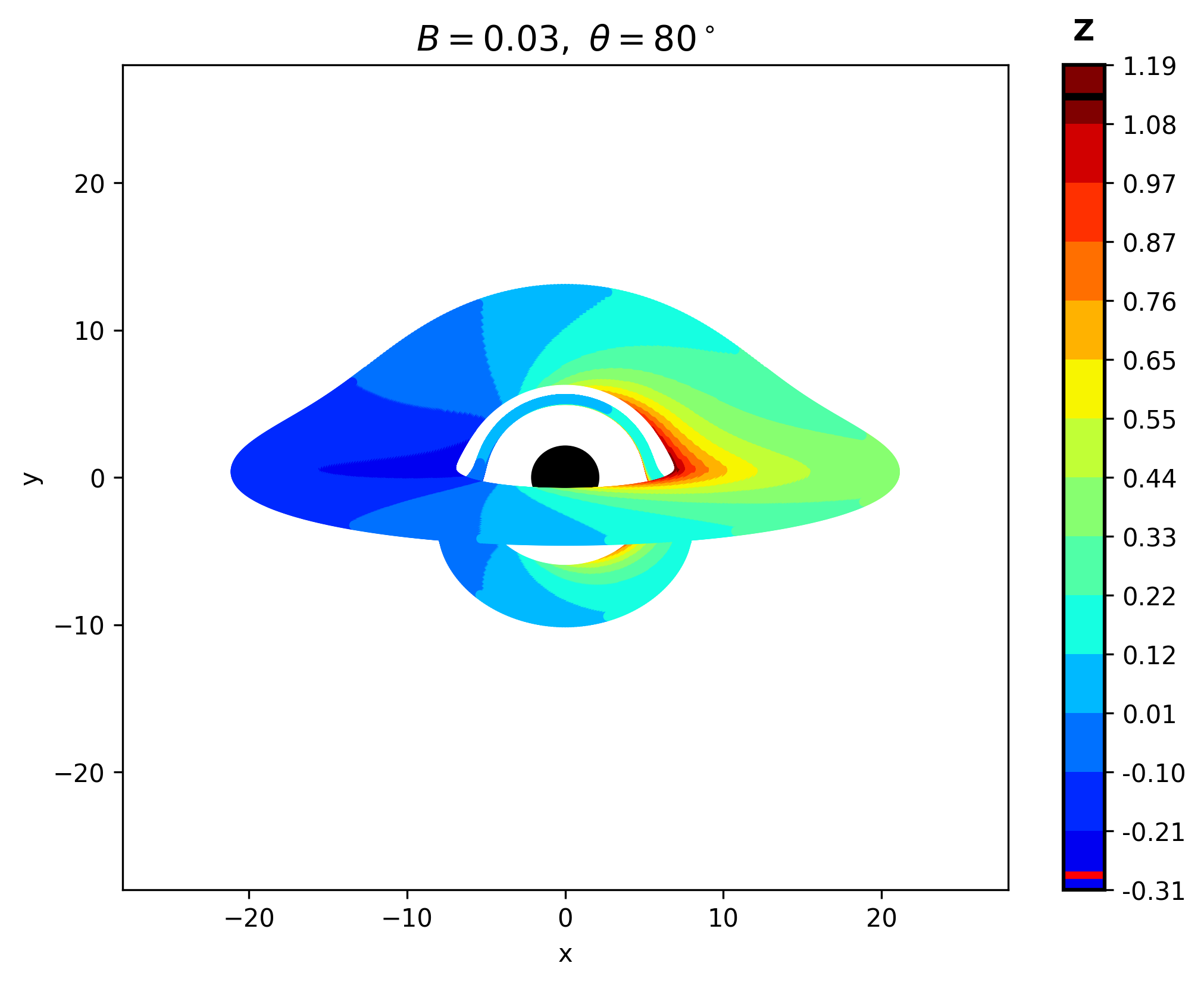}\hspace{-0.2cm}
 \includegraphics[scale=0.34]{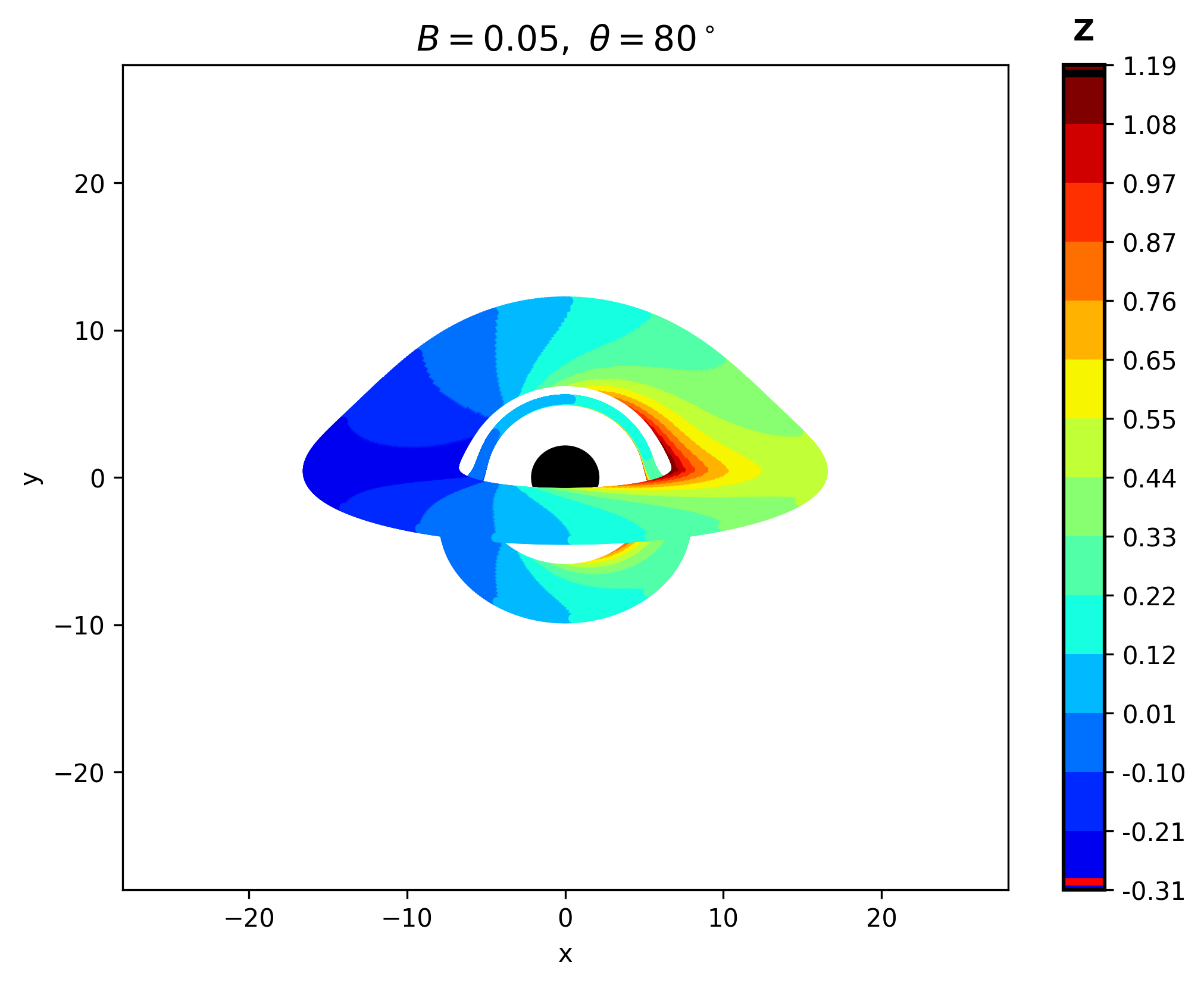}
 \end{tabular}
\caption{\label{fig:redeshiftdist}Direct and secondary images of the redshift factor $z$ distribution  of the accretion disk around the SBR BH  for different values of the inclination angle $\theta_0$ and magnetic field $B$.}
   \end{figure*}
\begin{figure*}
\begin{tabular}{ccc}
  \includegraphics[scale=0.35]{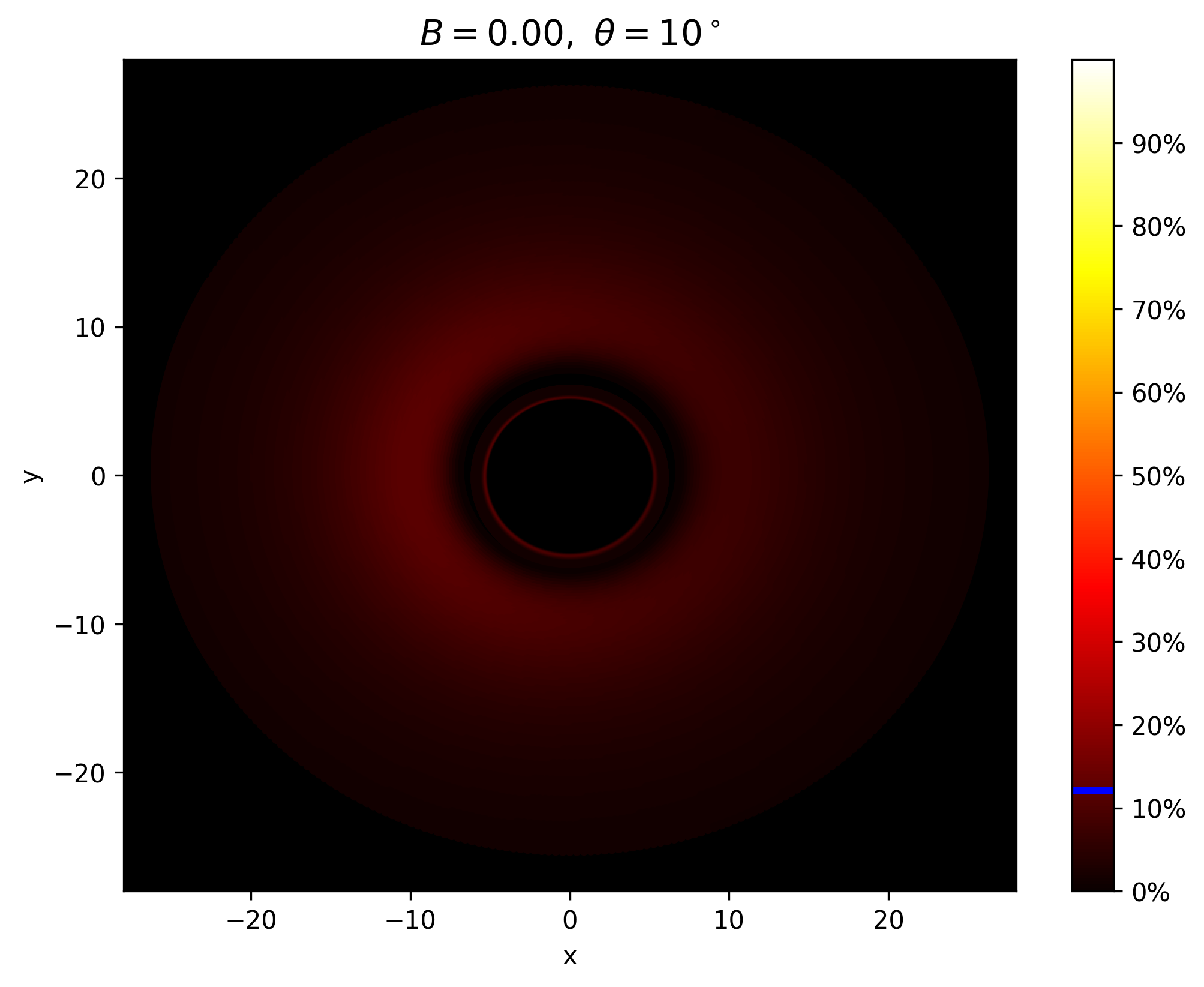}\hspace{-0.2cm}
   \includegraphics[scale=0.35]{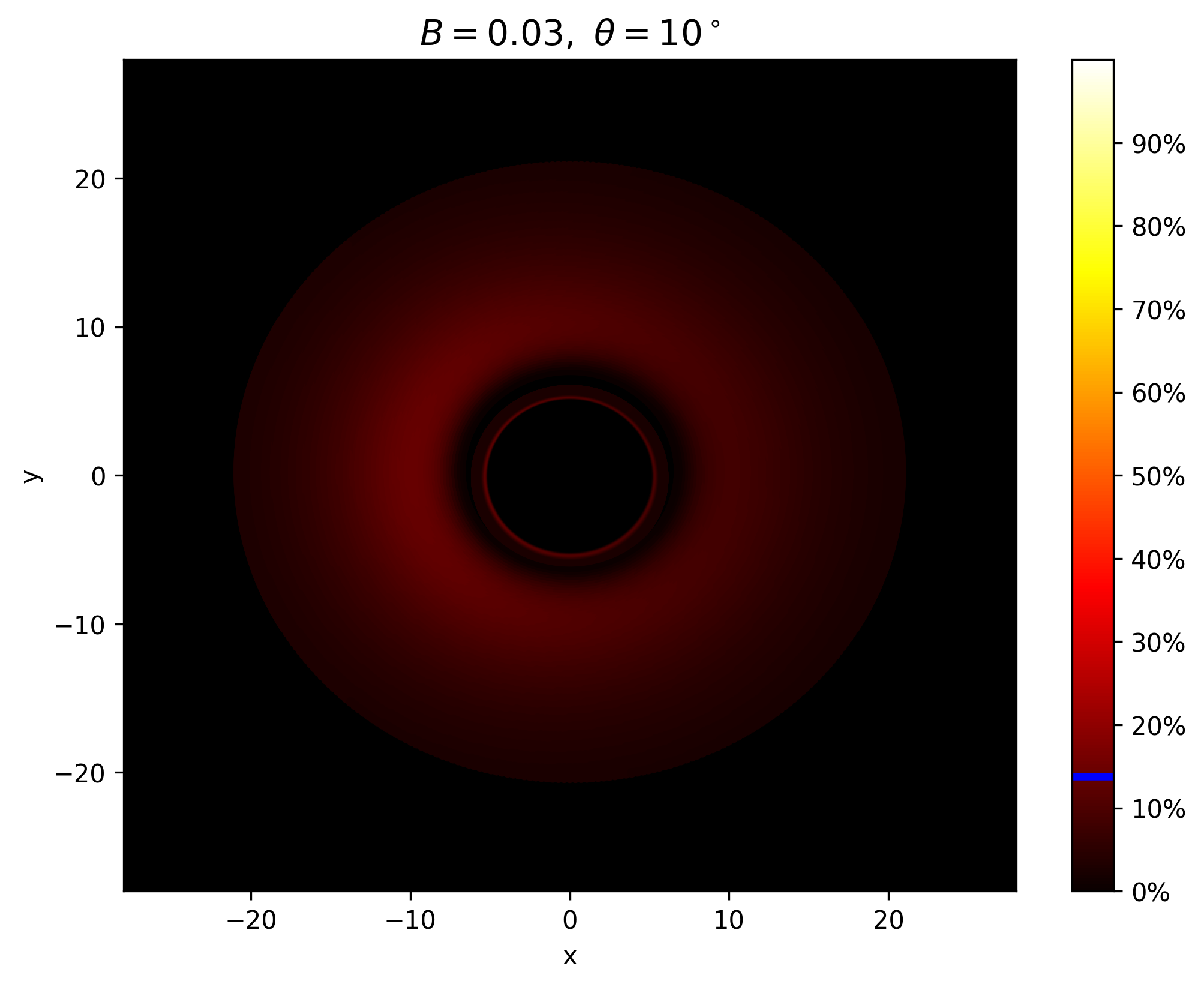}\hspace{-0.2cm}
  \includegraphics[scale=0.35]{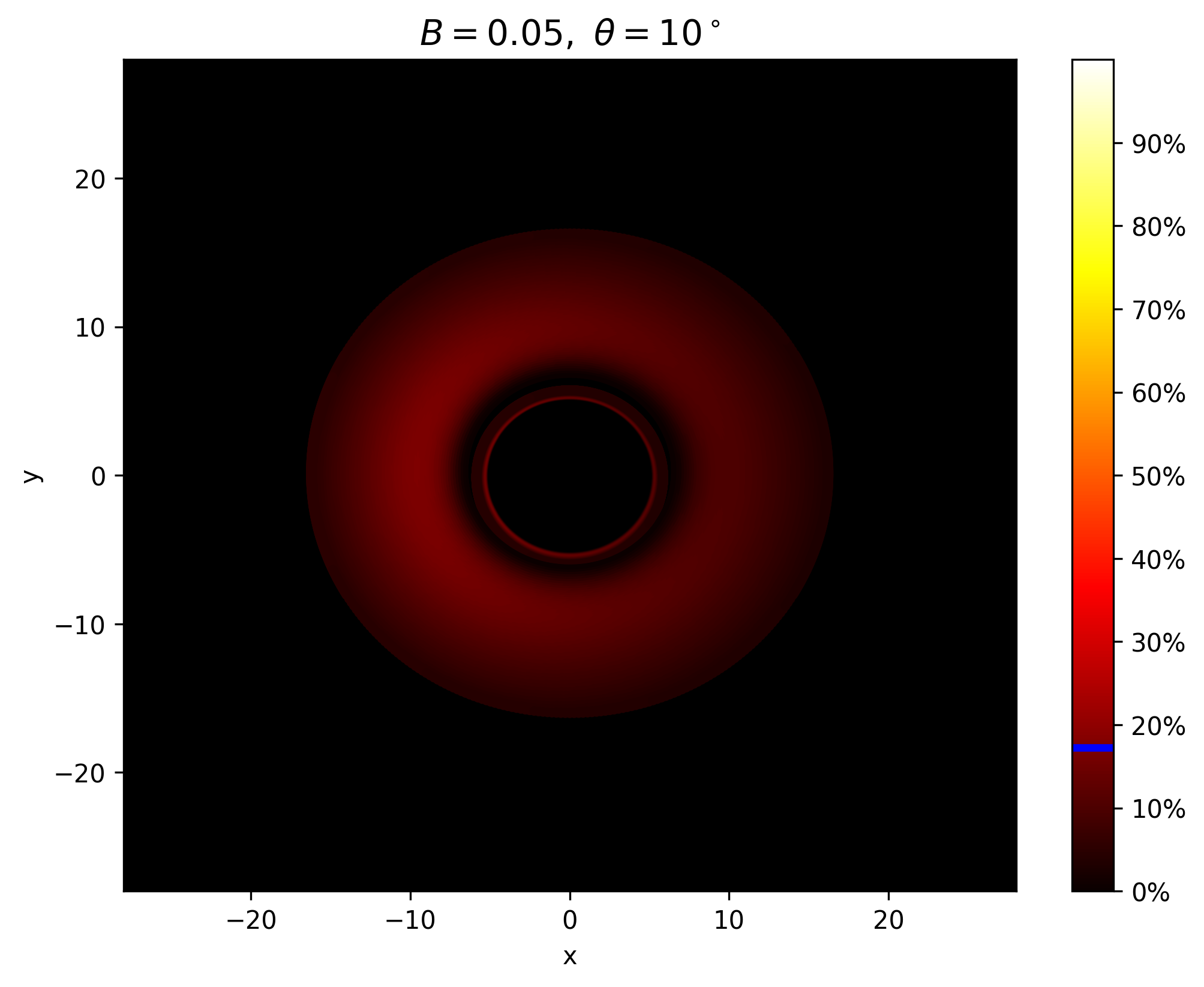}\\
  \includegraphics[scale=0.35]{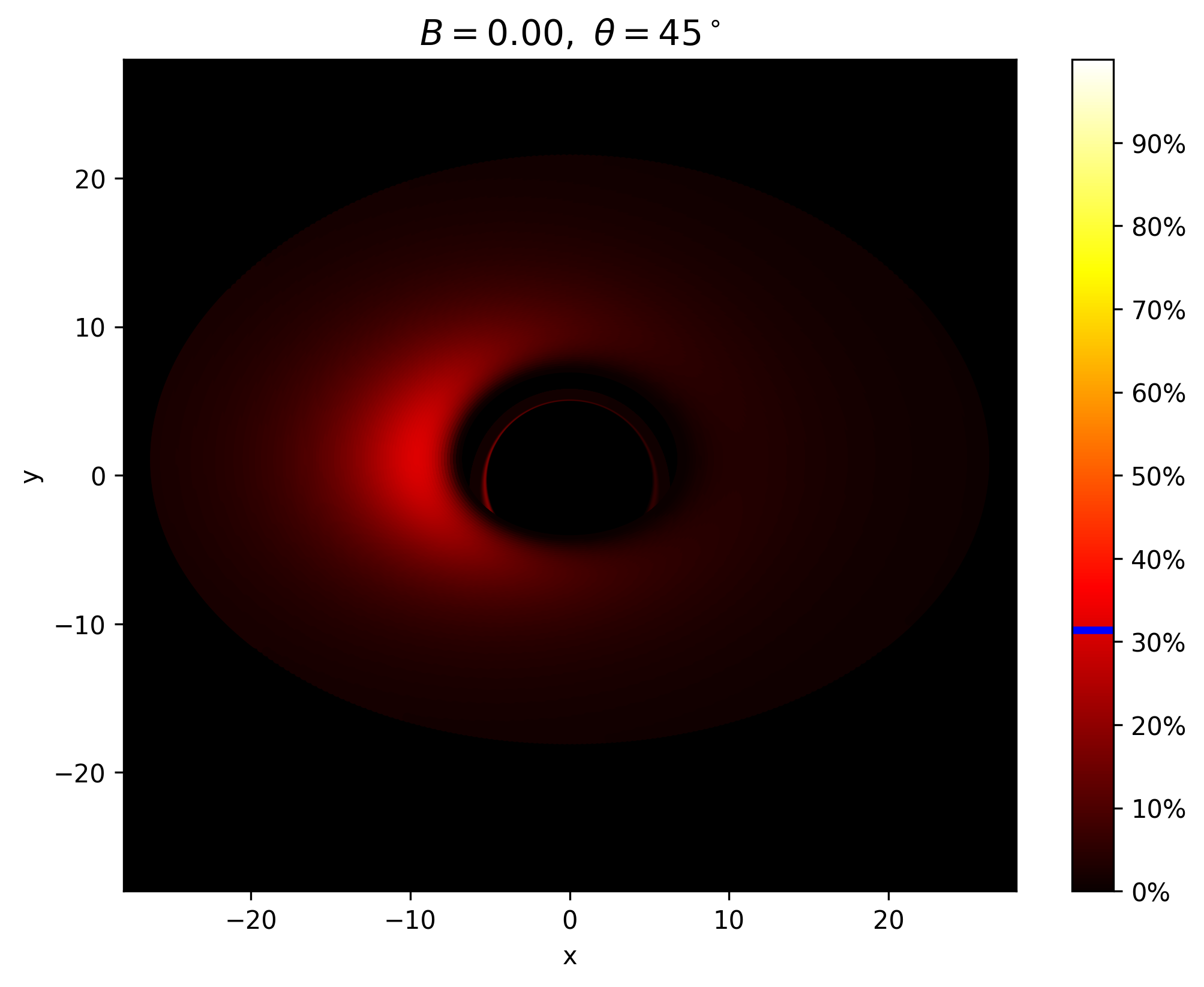}\hspace{-0.2cm}
  \includegraphics[scale=0.35]{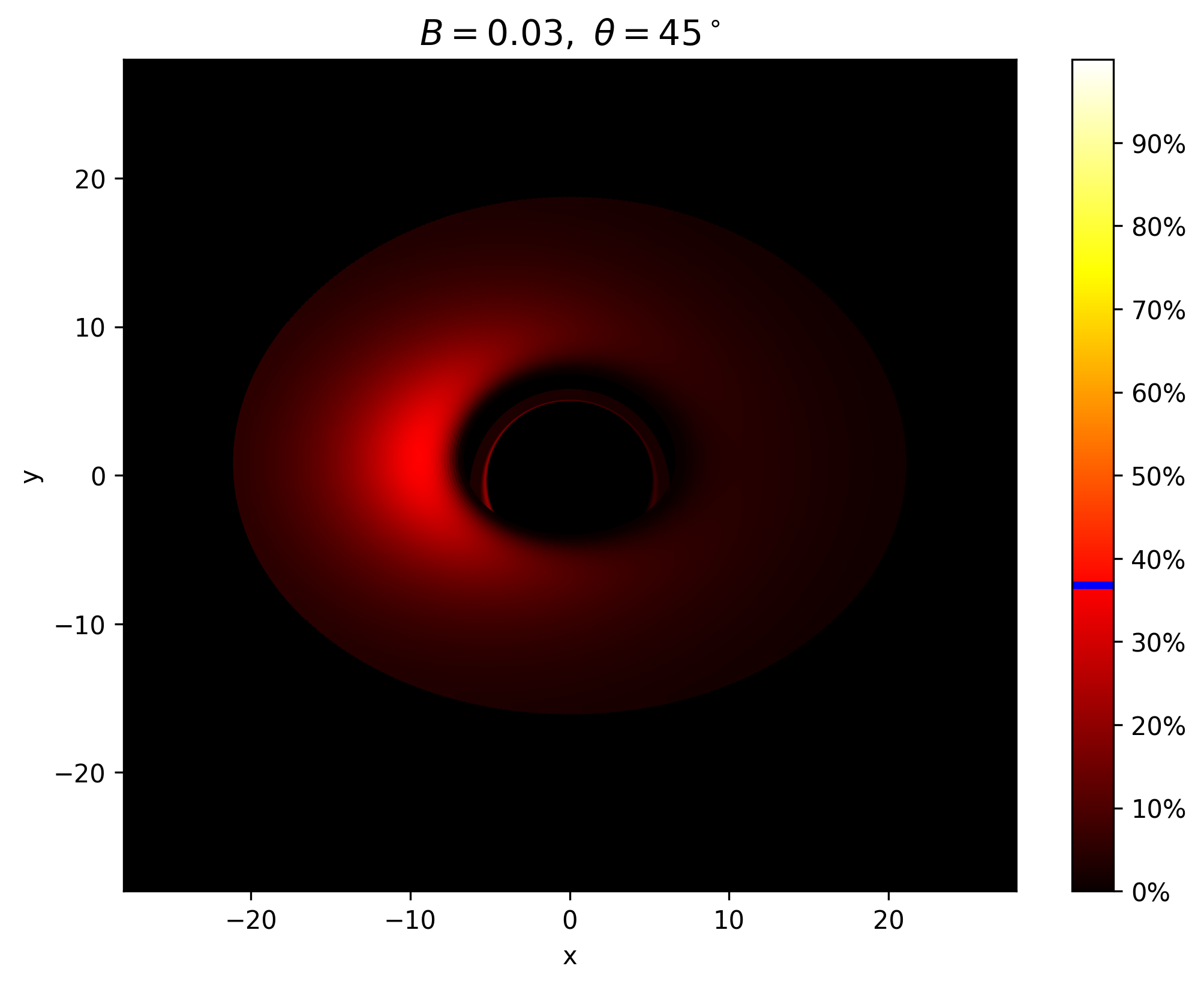}\hspace{-0.2cm}
  \includegraphics[scale=0.35]{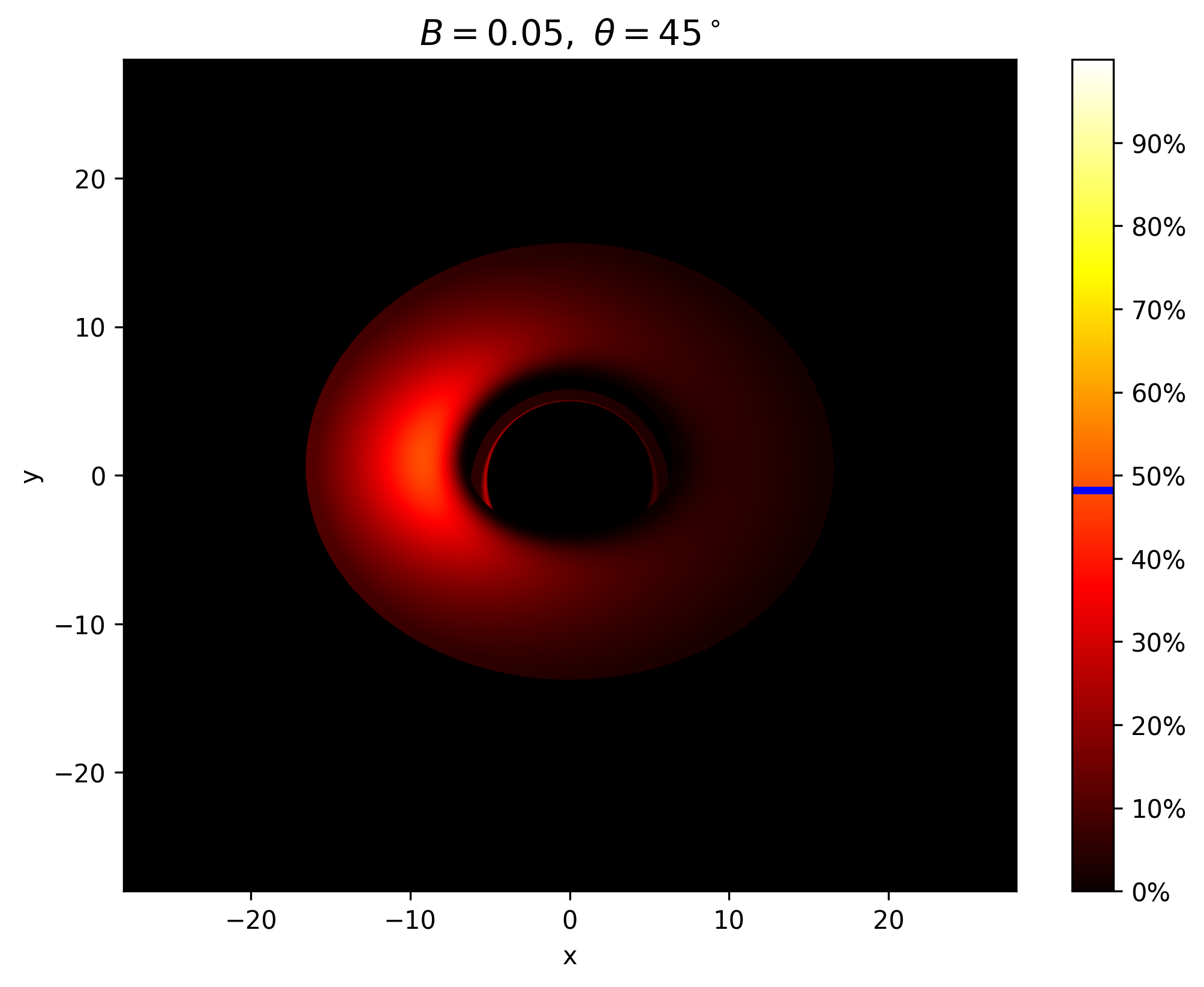}\\
  \includegraphics[scale=0.35]{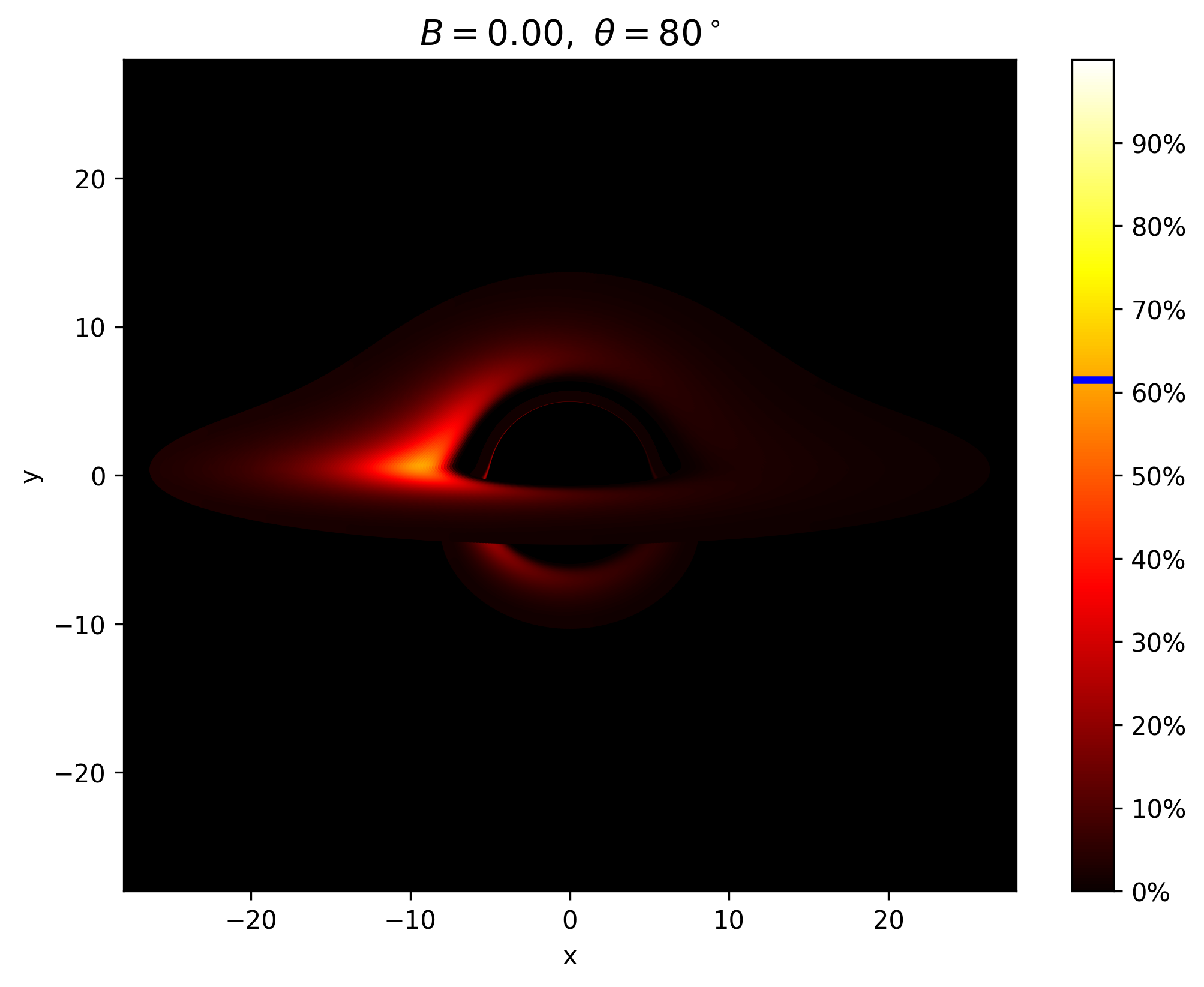}\hspace{-0.2cm}
  \includegraphics[scale=0.35]{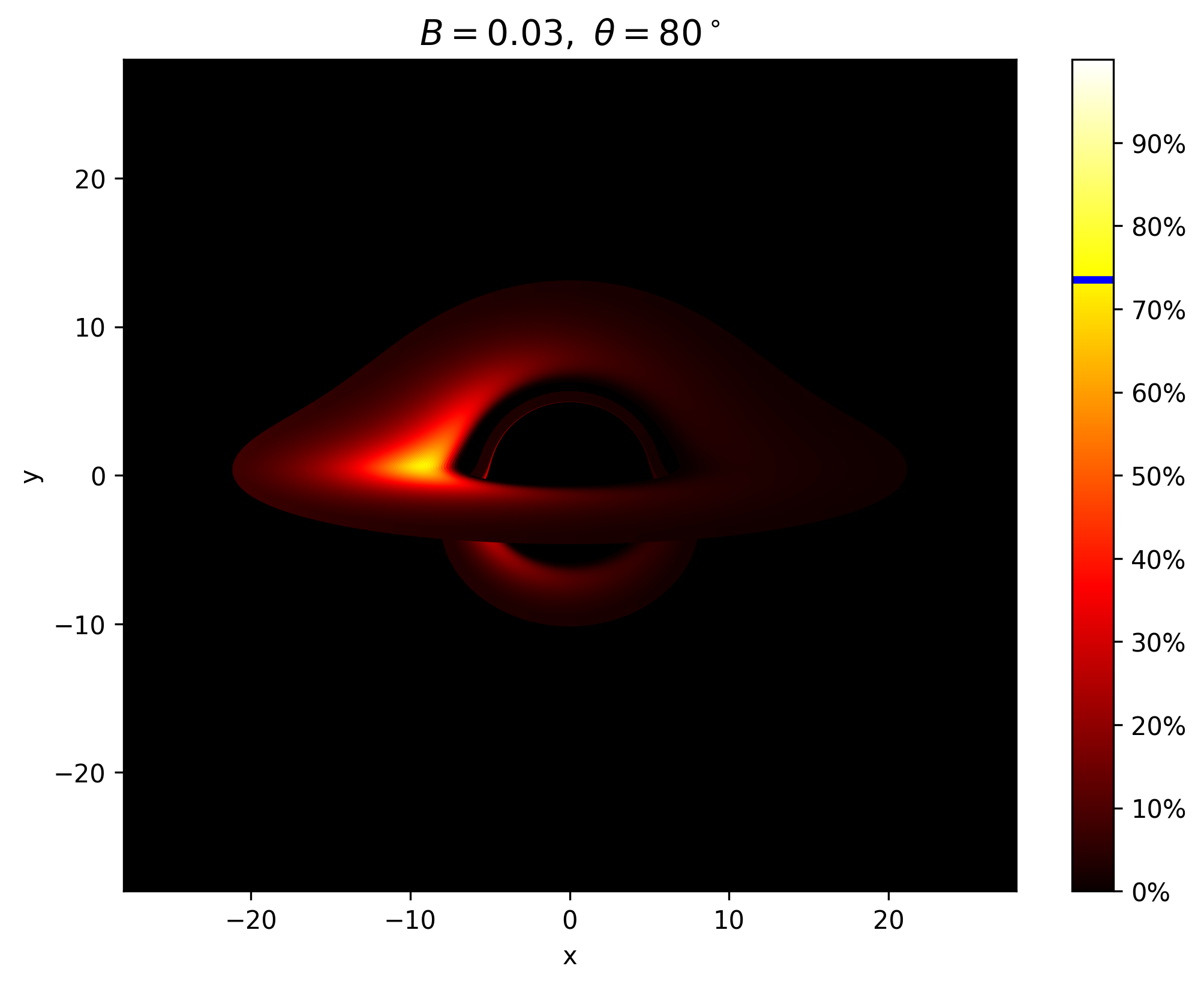}\hspace{-0.2cm}
  \includegraphics[scale=0.35]{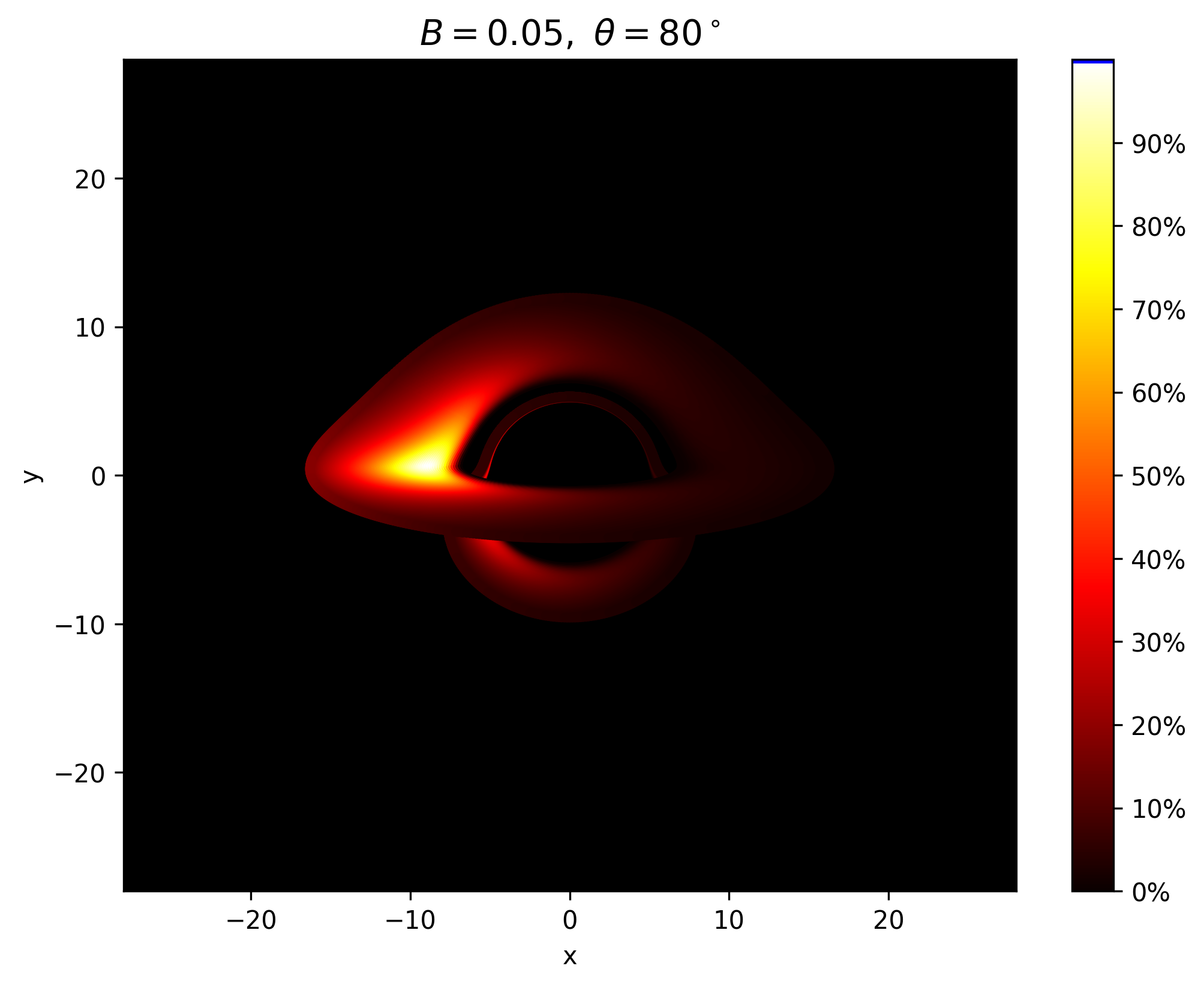}
  \end{tabular}
	\caption{\label{fig:fluxobs} Direct and secondary images of the observed flux $F_{obs}$ distribution  of the accretion disk around the SBR BH  for different values of the inclination angle $\theta_0$ and magnetic field $B$.}
\end{figure*}
\begin{figure*}
    \centering
    \includegraphics[width=0.48\linewidth]{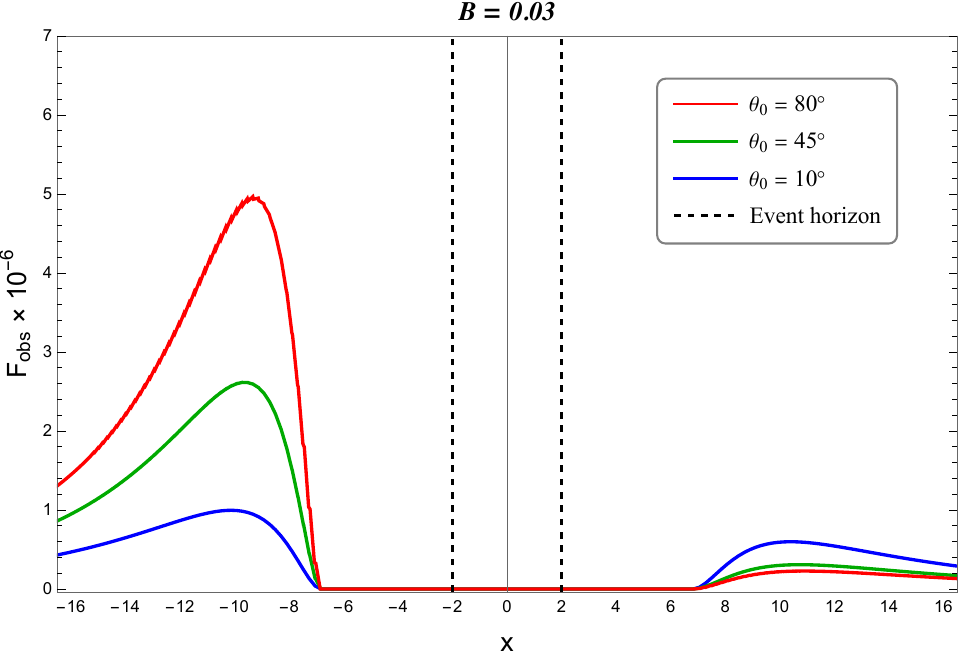}\hspace{0.5cm}
 \includegraphics[width=0.48\linewidth]{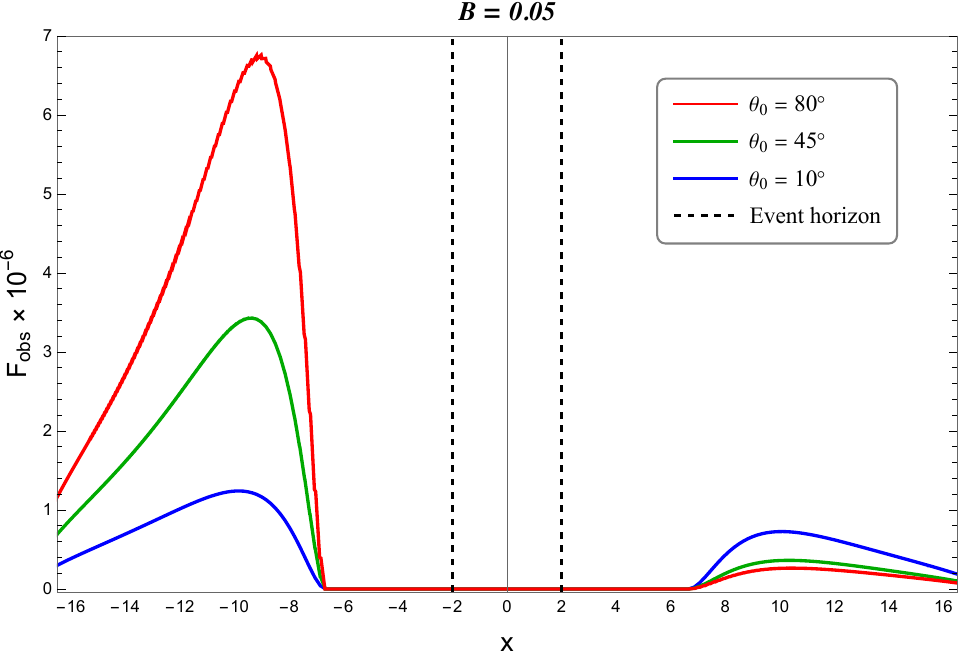}
    \caption{Observed flux $F_{obs}$ along the $y=0$ line as a function of $x$ in the camera frame. We fixed $B = 0.03$ (left panel), and $B = 0.05$ (right panel).}
    \label{fig:placeholder}
\end{figure*}

\subsection{Radiation Properties of Accretion Disks}\label{subs:B}
Based on the relativistic model, the radiation flux from the accretion disk is expressed as~\cite{1974ApJ...191..507T,Shakura:1972te,1973blho.conf..343N,Alloqulov:2024cto}
\begin{eqnarray}
F(r) = \frac{-\dot{M}_0 \Omega_{K ,r}}{4\pi \sqrt{-g}(E - \Omega_K L)^2}
\int_{r_{\text{isco}}}^{r} (E - \Omega_K L)L_{,r}dr\,. 
\label{eq:flux}
\end{eqnarray}
Here, $\dot{M}_0$ characterizes the mass accretion rate, and $g$ corresponds to the determinant of the metric. 

We derive the radiation temperature $T(r)$ of the disk utilizing the Stefan–Boltzmann law 
\begin{equation}
    F(r)=\sigma_{SB}T^4(r)\ .
    \label{eq:T}
\end{equation}

Fig.~\ref{fig:Flux}  illustrates the energy flux $F(r)$ (left) and radiation temperature $T(r)$ (right) in relation to the disk radius. Both plots start from the $R_{\text{ ISCO}}$ radius and extend outward. As $B$ increases, this flux and temperature distribution curves shift to higher values.  Also, in Fig.~\ref{fig:F_B}, the dependence of energy flux on magnetic field strength $B$ is demonstrated for different disk radii. With increasing $B$, the energy flux grows to larger values. When the disk is affected by the magnetic field, it is observed to be brighter and have a higher temperature in comparison to the Schwarzschild case.

\subsection{Observed image and Redshift factor}\label{subs:c}
Observers do not measure the energy flux directly. They measure the observed energy flux $F_{obs}$ at infinity, which is affected by gravitational redshift and the Doppler effect~\cite{Ellis2009,Sharipov:2025yfw}
\begin{eqnarray}
F_{obs} = \frac{F(r)}{(1+z)^4} \, .
\end{eqnarray}
Here, $z$ denotes the redshift factor~\cite{Luminet1979,Misner:1973prb,Zafar:2025nho}:
\begin{eqnarray}
1 + z =
\frac{1 + \Omega_K\, b \sin \theta \cos\alpha}{\sqrt{-g_{tt} - g_{\phi\phi}\Omega_K^2}}\, . 
\end{eqnarray}
\begin{table}
\centering
\caption{Maximum values of flux energy, temperature and redshift factor (for $\theta_0=80^\circ$) together with Schwarzschild BH case.}
\label{tab:metric_comparison}
\begin{tabular}{@{}lcccc@{}}
\toprule
\textbf{} & \textbf{Energy flux} & \textbf{Temperature} & \textbf{Redshift factor} \\ & $F_{\text{max}}/F_{\text{Schw}}$ & $T_{\text{max}}/T_{\text{Schw}}$ & $z_{\max}/z_{\text{Schw}}$ \\\hline 
$B=0.00$ & $1.000$ & $1.000$ & $1.000$ \\\hline 
$B=0.03$ & $1.139$ & $1.033$ & $1.021$ \\\hline 
$B=0.05$ & $1.424$ & $1.092$ & $1.059$ \\\hline\hline
\end{tabular}
\end{table}

In Table~\ref{tab:metric_comparison}, the observable radiation quantities of the accretion disk relative to the Schwarzschild BH case ($B = 0$) are listed. As $B$ increases, All ratios shift to higher values. This means that the external magnetic field enhances the maximum values of the radiation quantities of the accretion disk in comparison to the Schwarzschild case.

We now obtain the redshift factor distribution and the accretion disk image around the SBR BH by calculating, for each point, its observed value at the camera plane. Fig.~\ref{fig:redeshiftdist} demonstrates the observed redshift distribution of the radiation from the accretion disk orbiting the SBR BH. This effect arises from gravitational redshift and Doppler shift. In all plots, we set the same range and the same discrete color bar to make comparison easier. Each plot has different maximum and minimum values of the redshift factor. We mark the maximum value with a black line and the minimum with a magenta line in the color bar. As illustrated in the figure, in the $(B = 0.05, \theta_0 = 80^\circ)$ case, we obtain the highest $z_{\max} = 1.17$ and the lowest $z_{\min} = -0.3$ values of redshift among these plots. Even these maximum and minimum results are higher and lower, respectively, in comparison with the Schwarzschild case $(B = 0)$.

Fig.~\ref{fig:fluxobs} demonstrates the direct and secondary images of the accretion disk obtained from the observed energy flux. The observed flux maximum value $(F_{\text{obs}})_{\max}=6.74*10^{-6}$ is taken from the $(B=0.05,\theta_0=80^\circ)$ case among these plots. Then we normalize $(F_{\text{obs}})_{\max}\rightarrow100\%$ and use it for all plots as a general maximum. Each image has a color bar on the right side. The blue line in the color bar denotes the maximum value for the related image. The maximum observed flux value reaches $62\%$ in the Schwarzschild BH $(B = 0)$ at $\theta_0 = 80^\circ$. From this, the observed flux appears dimmer without a magnetic field.

Assuming that we have an observed image of an accretion disk. We mark a horizontal line $(y = 0)$ in the middle of the image. Then we calculate the observed flux $F_{\text{obs}}$ lying on this line. As illustrated in Fig.~\ref{fig:placeholder}, we demonstrate the effect of the inclination angle $\theta_0$ on the observed flux along the $y = 0$ line in the camera plane, where we set $B = 0.03$ (left panel) and $B = 0.05$ (right panel). Because the accretion disk rotates and the Doppler shift affects it, the observed flux appears asymmetric. These pronounced differences enable for a detailed analysis of the properties of accretion disks in such systems.

\section{Summary and Conclusions}\label{Sec:conclusion}

In this work, we investigated the optical and radiative signatures of an accretion disk around the SBR BH, which corresponds to the non-rotating limit of the Kerr-Bertotti-Robinson solution of the Einstein-Maxwell equations. We first analyzed the properties of the SBR spacetime and then studied photon propagation, null geodesics, and ray-tracing techniques using the Lagrangian formalism.

Based on the ray-tracing results, a bundle of photons originating from spatial infinity undergoes expansion in the presence of an external $B$ field (see Fig.~\ref{fig:ray1}). This effect can be attributed to the modification of the initial conditions employed in solving the photon orbital equations, which are required for analyzing optical phenomena in the vicinity of BHs. In particular, these initial conditions differ from those in the Schwarzschild BH case due to the use of Melvin-like magnetized spacetime metrics.

The key quantities, such as the event horizon radius $r_h$, the photon-sphere radius $r_{ph}$ and ISCO radius $r_{\rm ISCO}$, were determined for a BH in the presence of a magnetic field. Our findings showed that these quantities increase as the magnetic field parameter $B$ increases, inferring a strengthening of the gravitational field. Furthermore, for $B=0.05$, the range of impact parameters corresponding to lensed photon trajectories (emissions) is shifted toward smaller values, $b\in(4.976,5.149)\cup(5.19,6.128)$, compared to the Schwarzschild case $B=0$, as seen in Table~\ref{tab:nb}.

Building on these numerical results, we further performed an analytical study of thin accretion disk dynamics in this magnetized spacetime. By deriving the modified Keplerian frequency $\Omega_K = \sqrt{M/r^3}(1+B^2 r^2)$ and the specific energy $E = (1-2M/r - B^2M^2)/\sqrt{1-3M/r - B^2M^2 - MB^2r}$ for circular orbits, we obtained the exact ISCO radius $r_{\text{ISCO}} = 6M/(1-B^2M^2)$. For weak magnetic fields, this expands to $r_{\text{ISCO}} = 6M(1+B^2M^2 + \cdots)$, confirming the outward shift observed numerically in Table~\ref{tab:nb}. The radiative efficiency $\eta = 1 - E(r_{\text{ISCO}})$ was computed. For $B=0$, $\eta \approx 0.057$, the familiar Schwarzschild value. When the $B$ field is turned on, the efficiency decreases and this decrease becomes more pronounced as the field strength increases. For a field strong enough to modify the geometry ($\beta = BM \sim 0.1$), the efficiency drops to approximately $0.0053$, corresponding to a decrease of about $0.052$ or roughly $91\%$ relative to the unmagnetized case. This significant reduction arises because the outward shift of the ISCO increases the binding energy substantially, meaning considerably less energy is radiated away as the matter plunges into the BH. While this dramatic decrease may seem surprising, it is a direct consequence of the exact magnetized spacetime geometry. The $B$ field, through its non-perturbative coupling to gravity, fundamentally alters the structure of circular orbits, pushing the ISCO outward to larger radii where the gravitational potential well is shallower. This prediction represents a clean analytic result of this exact solution.

On the basis of the optical inverse path law, we applied the ray-tracing method to investigate photon trajectories around the SBR BH. Using this approach, we constructed the image-formation diagram and computed the observed image of the accretion disk. We found that the $B$ field, being encoded in the spacetime geometry, primarily affects the direct image of the accretion disk, leading to a reduction in its apparent size. Within a relativistic model, we also derived key radiation quantities, namely the energy flux, radiation temperature and redshift factor, from emission at the disk surface, as illustrated in Figs.~\ref{fig:FBline1} and~\ref{fig:accrthin}. As the magnetic field parameter $B$ increases, the maximum values of the energy flux, radiation temperature, and the redshift factor increase relative to the Schwarzschild case (see Fig.~\ref{fig:Flux} and Table~\ref{tab:metric_comparison}). For instance, for $B=0.05$ and $\theta_0=80^\circ$, we obtained the redshift factor as $z_{\max}=1.17$ and $z_{\min}=-0.3$, while  for $B=0$ and $\theta_0=80^\circ$ (Schw BH), the corresponding values are $z_{\max}=1.11$ and $z_{\min}=-0.28$, as shown in Fig.~\ref{fig:redeshiftdist}. Finally, extending the analysis to a more general case, we computed the observed flux distribution and constructed the corresponding images for a distant observer. As shown in Fig.~\ref{fig:fluxobs}, increasing $B$ leads to brighter and more compact images.

Our findings provide potential observational discriminants for strong-field regimes in BH-magnetar systems and establish a basis for understanding the impact of self-consistent electromagnetic fields on BH observables.

\begin{acknowledgements}
PS acknowledges the support of the Anusandhan National Research Foundation (ANRF) under the Science and Engineering Research Board (SERB) Core Research Grant (Grant No.\ CRG/2023/008980). 
\end{acknowledgements}

\bibliographystyle{apsrev4-1}
\bibliography{ref2,refnew,ref22}

\end{document}